\documentclass[aps,prd,a4paper,longbibliography,reprint,superscriptaddress,floatfix,nofootinbib,amsmath,amssymb]{revtex4-1}
\usepackage{color,hhline,psfrag,rotating}
\usepackage{dcolumn}
\usepackage{bm}
\usepackage[utf8]{inputenc}
\usepackage{hyperref}
\usepackage{multirow}
\begin{document}
\title{Lattice simulations
with \boldmath$N_f=2+1$ improved Wilson fermions at a fixed strange
quark mass}
\author{Gunnar S.\ Bali}
\email{gunnar.bali@ur.de}
\affiliation{Institute for Theoretical Physics, University of Regensburg, 93040 Regensburg, Germany}
\affiliation{Tata Institute of Fundamental Research, Homi Bhabha Road, Mumbai 400005, India}
\author{Enno E.\ Scholz}
\email{enno.scholz@ur.de}
\affiliation{Institute for Theoretical Physics, University of Regensburg, 93040 Regensburg, Germany}
\author{Jakob Simeth}
\email{jakob.simeth@ur.de}
\affiliation{Institute for Theoretical Physics, University of Regensburg, 93040 Regensburg, Germany}
\author{Wolfgang S\"oldner}
\email{wolfgang.soeldner@ur.de}
\affiliation{Institute for Theoretical Physics, University of Regensburg, 93040 
Regensburg, Germany}
\collaboration{RQCD Collaboration}
\date{\today}
\begin{abstract}
The explicit breaking of chiral symmetry of
the Wilson fermion action results in additive quark mass renormalization.
Moreover, flavour singlet and
non-singlet scalar currents acquire different renormalization constants
with respect to continuum regularization schemes.
This complicates keeping the renormalized strange
quark mass fixed when varying the light quark mass in simulations
with $N_f=2+1$ sea quark flavours.
Here we present and validate our strategy within the CLS
(Coordinated Lattice Simulations) effort to achieve this in simulations
with non-perturbatively order-$a$ improved Wilson fermions.
We also determine various combinations of renormalization
constants and improvement coefficients.
\end{abstract}

\maketitle

\section{Introduction}
With the gradual removal and reduction of systematic sources of error,
including finite volume, unphysical quark mass and lattice
spacing effects, Lattice QCD simulations have gained prominence
in predicting non-perturbative matrix elements that are of
phenomenological importance. Present-day large scale simulations
employ a multitude of quark actions, namely, overlap and
domain wall actions, staggered actions, twisted mass Wilson actions
at maximal twist and Wilson actions.

On the one hand overlap and domain wall actions (with a large extent in
the fifth direction) have the theoretically most desirable properties,
including automatic order-$a$ improvement and an exact chiral symmetry at
non-vanishing values of the lattice spacing $a$.
On the other hand Wilson fermions are cheaper to simulate in comparison
and no
approximations such as the uncontrolled rooting of fermionic determinants
are required.
Furthermore, unlike in the staggered or twisted mass formulations, no taste or
unphysical isospin symmetry breaking takes place: Simulating
QCD with $N_f=2+1$ flavours, where we assume the light quarks to be
mass-degenerate, there is only one pion mass $M_{\pi}$ and one kaon mass
$M_K$.

While in the other fermion formulations mentioned above lattice effects are of
order $a^2$ for most matrix elements, the naive Wilson action has artefacts of order $a$
that need to be removed non-perturbatively, in order to improve
the action and operators.
Another draw-back is additive quark mass renormalization. While in the
continuum only for the axialvector current
a distinction between flavour singlet and non-singlet dimension three quark
bilinears needs to be made, in the Wilson formalism these
quark mass combinations (or, equivalently, scalar currents)
renormalize differently. In simulations with dynamical sea quarks
this complicates parameter tuning within the quark mass plane.

\begin{figure}
\includegraphics[width=0.49\textwidth]{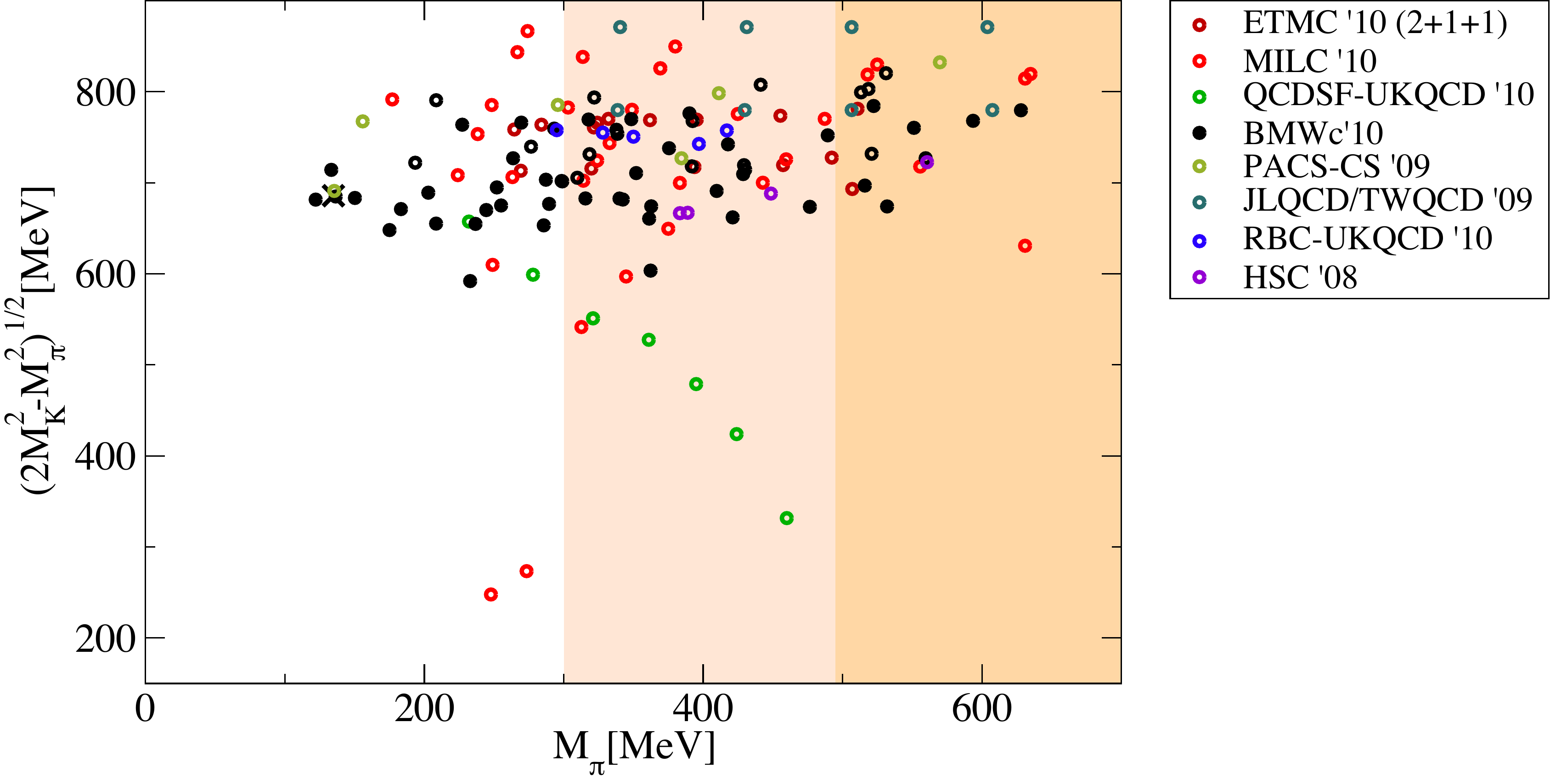}
\caption{Positions of some lattice simulations in the quark mass plane, taken from
Ref.~\protect\cite{Hoelbling:2014uea} ($M_{\pi}\sim \sqrt{m_{\ell}}$, $(2M_K^2-M_{\pi}^2)^{1/2}\sim \sqrt{m_s}$). For details, see also
Refs.~\protect\cite{Hoelbling:2011kk,Fodor:2012gf}.}
\label{fig:overview}
\end{figure}

Traditionally, simulations with $N_f=2+1$ (or $N_f=2+1+1$) fermions
have been performed, keeping the strange quark mass approximately fixed
while reducing the light quark mass towards its physical point value.
For an overview of simulation points of many lattice
collaborations that were available a few years ago, see Fig.~\ref{fig:overview},
taken from Ref.~\cite{Hoelbling:2014uea}. More details
can be found in the reviews \cite{Hoelbling:2011kk,Fodor:2012gf}. For recent simulations see, e.g., Refs.~\cite{Bietenholz:2010jr,Bietenholz:2011qq,Inoue:2011ai,Bruno:2014jqa,Blum:2014tka,Aoki:2015pba,Orginos:2015aya,Ishikawa:2015rho,Yoon:2016dij} ($N_f=2+1$) and \cite{Herdoiza:2013sla,Bazavov:2014wgs,Borsanyi:2014jba} ($N_f=2+1+1$). 
Determining the point in the quark mass plane where
the pion and kaon masses assume their physical values requires
knowledge of the lattice spacing. This can only be obtained by
extrapolating a third dimensionful observable to this physical point.
Once the scale is known, the initial guesses for the pion and
kaon masses may very well turn out to be incorrect.
Then new points, e.g., with a different
strange quark mass, need to be simulated, or quark mass reweighting
becomes necessary.

It was realized by the QCDSF collaboration~\cite{Bietenholz:2010jr}
that extrapolating along a line where the sum of lattice quark masses
$m_u+m_d+m_s=2m_{\ell}+m_s$ is kept constant
(the green outliers in Fig.~\ref{fig:overview}) can be very advantageous,
since gluonic observables
or the centres of mass of meson and baryon multiplets
dominantly depend on the trace of the quark mass matrix but
are only mildly affected by
the difference $m_s-m_{\ell}$.
This allows for a better controlled approach of the physical
point along this mass plane trajectory, with only a small variation of
the lattice scale, minimizing the risk of missing the physical
point. Therefore, only
a very moderate subsequent reweighting of quark masses --- if at
all --- may become necessary.
Within the $N_f=2+1$ CLS effort~\cite{Bruno:2014jqa} we follow this strategy.

At moderately fine lattice spacings, $a\gtrsim(3\,\textmd{GeV})^{-1}$, physical
point simulations with a spatial extent of the lattice
$L>4/M_{\pi}$ are possible. However, for complicated observables
or at finer lattice spacings, obtaining meaningful physical point results
is still prohibitively expensive. In this situation it is desirable
to have a second line in the
quark mass plane at hand to validate any extrapolation strategy.
This has motivated us within the CLS effort to generate ensembles not only at fixed values of
$2m_{\ell}+m_s$ but also to set the renormalized strange
quark mass (close) to its physical value while lowering the
light quark mass. This additional trajectory allows us to
benefit from $\textmd{SU}(2)$ chiral perturbation theory (ChPT)
and to determine the corresponding low energy constants.
$\textmd{SU}(2)$ ChPT should be more
reliable than $\textmd{SU}(3)$ ChPT as the
$K$ and the (hypothetical) octet $\eta_8$ mesons are not particularly 
light in nature or in the simulations envisaged here. The additional
line in the quark mass plane also provides an alternative to
Gell-Mann--Okubo style expansions in the $\textmd{SU}(3)$
flavour symmetry breaking
parameter~\cite{GellMann:1962xb,Okubo:1961jc,Bietenholz:2011qq}.

Due to the different renormalization patterns of singlet and
non-singlet quark mass combinations, it is not straightforward
to keep the renormalized strange quark mass fixed in simulations
with Wilson fermions. This is evident from Fig.~\ref{fig:overview}
where no single group managed to keep the renormalized strange quark mass
constant, illustrating that
even when employing other fermion actions the correct tuning may not
be entirely trivial. Here we describe how we achieved an almost
constant renormalized strange quark mass,
starting from a few existing simulation points with
$2m_{\ell}+m_s= 3 m_{\mathrm{symm}} =\mathrm{const}.$
and additional points along the $m_s=m_{\ell}$ line that
usually will exist, either from searches for
the starting point of the $2m_{\ell}+m_s=3m_{\mathrm{symm}}$
trajectory or from non-perturbative renormalization efforts.

This article is organized as follows.
In Sec.~\ref{sec:method} we introduce our notations
and describe the basic method to achieve a fixed strange quark mass
in our simulations, including order-$a$ improvement.
In Sec.~\ref{sec:implement} we give a brief overview of
the action used and ensembles generated. We then parameterize
our quark mass data and also fit previously undetermined
improvement coefficients.
In Sec.~\ref{sec:mass} we
describe our determination of the physical quark mass point.
More details on the numerical results and fits are presented
in Sec.~\ref{sec:test}, where we also discuss combinations of
improvement coefficients, before we conclude in the final section.

\section{The method}
\label{sec:method}
We define our notation and derive useful relations,
which enable us to relate the strange quark
and light quark hopping parameters, defining a line of an (almost)
constant renormalized strange quark mass. We then
order-$a$ improve this result and discuss how to
keep constant renormalized masses of additional valence
quark flavours.
\subsection{Definitions and useful relations}
\label{sec:renorm}
We closely follow the notation of Ref.~\cite{Bhattacharya:2005rb},
however, substituting $\mathrm{Tr}\,M=3\overline{m}$, see below. We
assume two mass-degenerate flavours of light sea quarks
with masses $m_1=m_2=m_{\ell}$  and one strange sea quark ($m_3=m_s$).
We define $\kappa_{\mathrm{crit}}$ as the hopping parameter
value at which the quark mass from the axial Ward-Takahashi identity (AWI mass)
in the flavour-symmetric case $m_1=m_2=m_3$
vanishes. Lattice quark masses are then defined as\footnote{Note that,
away from the $m_s=m_{\ell}$ line, $m_{\ell}$ defined in this way
can be negative for positive AWI masses. The average mass $\overline{m}$
remains positive.}
\begin{equation}
m_j=\frac{1}{2a}\left(\frac{1}{\kappa_j}-\frac{1}{\kappa_{\mathrm{crit}}}\right)
\,.
\end{equation}
We introduce the following conventions for averages:
\begin{align}
m_{jk}&\equiv\frac12(m_j+m_k)\,,\\
\overline{m}&\equiv\frac13(m_s+2m_{\ell})\,,\\
\overline{m^2}&\equiv\frac13(m_s^2+2m_{\ell}^2)\,.
\end{align}

We consider flavour non-singlet ($j\neq k$) pseudoscalar
$P^{jk}=\bar{q}^j\gamma_5q^k$ and axialvector
$A^{jk,0}_{\mu}=\bar{q}^j\gamma_{\mu}\gamma_5q^k$ currents.
The pseudoscalar current is automatically order-$a$ improved
while the improved axial current reads $A_{\mu}^{jk}=
A_{\mu}^{jk,0}+ac_A\partial_{\mu}P^{jk}$,
where $\partial_{\mu}$ is the symmetrized discrete next neighbour
derivative and the improvement coefficient $c_A$ was determined
non-perturbatively in
Ref.~\cite{Bulava:2015bxa}.
We define renormalized, order-$a$ improved currents
\begin{align}
\label{eq:renorm2}
\widehat{A}_{\mu}^{jk}&=Z_AA_{\mu}^{jk}\left[1+a(3\bar{b}_A\overline{m}+
b_Am_{jk})\right]\,,\\
\widehat{P}^{jk}&=Z_PP^{jk}\left[1+a(3\bar{b}_P\overline{m}+
b_Pm_{jk})\right]\,.
\end{align}
Note that $Z_P$ will depend on the target
renormalization scheme and scale. The factors
of 3 are due to the conventions used for $N_f=3$ in
Ref.~\cite{Bhattacharya:2005rb}. For the action we use,
$Z_A$ was calculated in Ref.~\cite{Bulava:2016ktf}.
Renormalized quark masses can be obtained
from the axial Ward identity (AWI)
\begin{equation}
\label{eq:quarkm}
\widehat{m}_j+\widehat{m}_k=2\widehat{m}_{jk}=
\frac{\partial_4\langle 0|\widehat{A}_4^{jk}|\pi^{jk}\rangle}
{\langle 0|\widehat{P}^{jk}|\pi^{jk}\rangle}\,,
\end{equation}
where $\pi^{jk}$ is a pseudoscalar
state with quark $q^j$ and antiquark $\bar{q}^k$.

Finally, we define unrenormalized (but improved) non-singlet
AWI masses:
\begin{equation}
\widetilde{m}_{jk}=
\frac{\partial_4\langle 0|A_4^{jk}|\pi^{jk}\rangle}
{2\langle 0|P^{jk}|\pi^{jk}\rangle}\,.
\end{equation}
These can easily be related to the renormalized quark masses
via Eqs.~\eqref{eq:renorm2}--\eqref{eq:quarkm}.
The main CLS ensembles~\cite{Bruno:2014jqa} are 
generated along
trajectories of constant average lattice quark masses $\overline{m}$
(and therefore $\overline{\widehat{m}}$ is constant up to $\mathcal{O}(a)$
corrections). For non-perturbative renormalization
purposes we generated additional ensembles
along the $\textmd{SU}(3)$ flavour symmetric
trajectory, i.e.\ $m_s=m_{\ell}$.
We now wish to keep $\widehat{m}_s$ fixed, varying $m_{\ell}$
and adjusting $\kappa_s$ as required.

The renormalized quark masses can be related
to the lattice quark masses:
\begin{align}
\widehat{m}_j&=Z_m\left\{m_j+(r_m-1)\overline{m}+a
\Bigl[b_mm_j^2+3\bar{b}_mm_j\overline{m}\right.\nonumber\\
&\qquad\left.\left.+3(r_m\bar{d}_m-\bar{b}_m)\overline{m}^2
+(r_md_m-b_m)\overline{m^2}
\right]\right\}\,.
\label{eq:massr}
\end{align}
At an average quark mass $\overline{m}>0$ the
coupling $g^2=6/\beta$ that corresponds to a given lattice spacing
$a$ will undergo renormalization too,
$g^2\mapsto \tilde{g}^2=g^2(1+b_ga\overline{m})$, 
where $b_g=0.012000(2)N_fg^2+\mathcal{O}(g^4)$~\cite{Luscher:1996sc}.
This means that $Z_J=Z_J[\tilde{g}^2,a(\tilde{g}^2)\mu]$,
where we consider $J\in\{m,P,A\}$,
and $r_m=r_m[\tilde{g}^2,a(\tilde{g}^2)\mu]$. The
dependence on the renormalization scale $\mu$ is absent in
the case of $Z_A$, as the non-singlet axial current does not carry an
anomalous dimension. 
Also the order-$a$ improvement coefficients
are functions of $\tilde{g}^2$, however,
in these cases we can neglect the effect of the difference
between $g^2$ and $\tilde{g}^2$, which is of a higher order in $a$.

Expanding $Z_J$ around $g^2$ gives
\begin{widetext}
\begin{align}
\nonumber
Z_J\left[\tilde{g}^2,a(\tilde{g}^2)\mu\right]&=
Z_J\left[g^2,a(g^2)\mu\right] \left\{
1+\left[\frac{\partial \ln Z_J(g^2,a\mu)}{\partial g^2}+
\frac{\mathrm{d} \ln Z_J(g^2,a\mu)}{\mathrm{d} \ln a}\frac{\partial\ln a(g^2)}{\partial{g^2}}\right]g^2b_ga\overline{m}+\ldots\right\}\\
&=
Z_J\left[g^2,a(g^2)\mu\right]\left\{1+\left[\frac{\partial\ln{Z_J(g^2,a\mu)}}{\partial g^2}-\frac{\gamma_J(g^2)}{4\pi\beta(g^2)}\right] g^2b_ga\overline{m}\right\}\,,
\end{align}
\end{widetext}
and similarly for $r_m$. Note that in this case
$\mathrm{d}\ln r_m/\mathrm{d}\ln a=0$. The same holds for the scale
dependence of $Z_A$ while $Z_m$ carries an anomalous dimension.
Above, we have introduced the
$\beta$-function
\begin{equation}
\beta(g^2)=-\frac{1}{4\pi}\frac{\mathrm{d}g^2}{\mathrm{d}\ln a}=
-\frac{g^2}{2\pi}\left(\beta_0\frac{g^2}{16\pi^2}
+\cdots\right)\,,
\end{equation}
where our normalization convention corresponds to
$\beta_0=11-\frac23N_f$, and the anomalous dimension of the current 
(or quark mass) $J$, which reads
\begin{equation}
\gamma_J(g^2)=\frac{\mathrm{d} \ln Z_J}{\mathrm{d} \ln a}\,.
\end{equation}

We can eliminate $b_g$ by redefining
\begin{align}
\label{eq:redef}
\tilde{b}_J(g^2)&\equiv
\bar{b}_J(g^2)\\\nonumber&+\frac{b_g(g^2)}{N_f}
\left[\frac{\partial\ln Z_J(g^2,a\mu)}{\partial g^2}
-\frac{\gamma_J(g^2)}{4\pi\beta(g^2)}\right]g^2\,.
\end{align}
Analogously, the $\bar{d}_m$ improvement coefficient
can be replaced by $\tilde{d}_m$ to
absorb the effect of $b_g$ on $r_m$.
Note that the anomalous dimension cancels from the above combination.
Both $\bar{b}_J$ and
$\tilde{b}_J$ are of $\mathcal{O}(g^4)$ in perturbation theory
and at present unknown.
With these substitutions Eq.~\eqref{eq:massr} becomes
\begin{align}
\widehat{m}_j&=Z_m\left\{m_j+(r_m-1)\overline{m}+a
\left[b_mm_j^2+3\tilde{b}_mm_j\overline{m}\right.\right.\nonumber\\
&\qquad\left.\left.+3(r_m\tilde{d}_m-\tilde{b}_m)\overline{m}^2
+(r_md_m-b_m)\overline{m^2}
\right]\right\}\,,
\label{eq:massr3}
\end{align}
where $Z_m$ now is a function of $g^2$, rather than of $\tilde{g}^2$.
The only other difference between the two equations is the
replacement of the bar-coefficients by tilde-coefficients.

Equation~\eqref{eq:massr3} implies that we can re-express the
(unrenormalized) AWI masses in terms of
the lattice masses, see Eqs.~(48)--(53) of
Ref.~\cite{Bhattacharya:2005rb}:
\begin{align}
\widetilde{m}_{jk}&=
\frac{Z_PZ_m}{Z_A}\Bigl[m_{jk}+(r_m-1)\overline{m}\nonumber\\&\quad+\left.
a\left(
\mathcal{A}m_{jk}^2+3\mathcal{B}m_{jk}\overline{m}
+9\mathcal{C}\overline{m}^2+3\mathcal{D}\overline{m^2}\right)\right]\,,
\label{eq:sharpe}
\end{align}
where
\begin{align}
\label{eq:adef}
\mathcal{A}&=b_P-b_A-2b_m\,,\\
\mathcal{B}&=\tilde{b}_P-\tilde{b}_A+\tilde{b}_m+2b_m+\frac13(r_m-1)(b_P-b_A)\,,\\
\mathcal{C}&=\frac13\left[(r_m-1)(\tilde{b}_P-\tilde{b}_A)+r_m\tilde{d}_m-\tilde{b}_m\right]
-\frac{b_m}{2}\,,\\
\mathcal{D}&=\frac13\left(r_md_m+\frac{b_m}{2}\right)\,.
\end{align}
Note that we independently
verified these results of Ref.~\cite{Bhattacharya:2005rb}.
However, we replaced $\bar{b}_J\mapsto \tilde{b}_J$ and
$\bar{d}_m\mapsto\tilde{d}_m$, absorbing the effect of
$b_g$ into these new coefficients. The difference between these
two sets of improvement coefficients is of $\mathcal{O}(g^4)$ and
is given in Eq.~\eqref{eq:redef}.
The relation between $N_f=3$ quark masses
$m_j^2+m_k^2=-4m_{jk}^2+12m_{jk}\overline{m}-9\overline{m}^2+3\overline{m^2}$
was used to derive Eq.~\eqref{eq:sharpe}.
For mass-degenerate up and down quarks
it will turn out more convenient to rewrite the order-$a$ corrections
in terms of their functional dependence on $(m_s-m_{\ell})^2$,
$(m_s-m_{\ell})\overline{m}$ and $\overline{m}^2$.

Some of the above improvement coefficients as well as
$Z_m$, $Z_P$ and $Z_A$ have been
computed to one-loop order in perturbation theory~\cite{Taniguchi:1998pf}
for the tree-level Symanzik improved gauge action
that we use in our simulations:\footnote{For the $Z_J$ we quote
the more precise values given in Ref.~\cite{Constantinou:2014fka},
using $c_{SW}=1$, which is consistent to this order.
The ordering of the numerical values in the first line
corresponds to $Z_m=Z_S^{-1}$, $Z_P$ and $Z_A^{-1}$, where
$Z_m$ and $Z_P$ are the conversions to the $\overline{\mathrm{MS}}$ scheme
at the scale $\mu=a^{-1}$. The scale and scheme dependence
cancels from the combination $Z$.}
\begin{align}
\label{eq:perturb}
Z&\equiv\frac{Z_mZ_P}{Z_A}
=1+(0.09546-0.11058+0.06786)C_Fg^2\nonumber\\
&=1+0.05274C_Fg^2\,,\\
b_m&=d_m=-\frac12-0.05722(5)C_Fg^2\,,\label{eq:perturb1}\\
b_A&=1+0.0881(1)C_Fg^2\,,\label{eq:perturb2}\\
b_P&=1+0.0890(1)C_Fg^2\label{eq:perturb3}\,,
\end{align}
where $C_F=4/3$ and terms of $\mathcal{O}(g^4)$ are ignored.
This means the difference
$b_A-b_P$ practically vanishes to $\mathcal{O}(g^2)$
while $\tilde{b}_A-\tilde{b}_P$, being a sea
quark effect, exactly vanishes to this order.
Therefore, keeping the renormalized strange quark mass
$\widehat{m}_s$ constant amounts to keeping the AWI mass
\begin{align}
\widetilde{m}_s&\equiv 2\widetilde{m}_{13}-\widetilde{m}_{12}\nonumber\\
&=\widehat{m}_s\frac{Z_P}{Z_A}\left\{1+a\left[3(\tilde{b}_P-\tilde{b}_A)\overline{m}+(b_P-b_A)m_s\right]\right\}\label{eq:hatmass}
\end{align}
fixed, up to $\mathcal{O}(g^4 a\overline{m})$ and
$\mathcal{O}(g^2 am_s)$ corrections with small coefficients.
It has been confirmed non-perturbatively~\cite{Korcyl:2016ugy} that the
$am_s$ term is not unnaturally large.
In the absence of other information from non-perturbative approaches,
in particular on
$\tilde{b}_A-\tilde{b}_P$, we will keep $\widetilde{m}_s$ fixed instead of
$\widehat{m}_s$.

We remark that for the gauge and fermion action that we use
$r_m$ has been computed
perturbatively~\cite{Constantinou:2014rka},
with the result
\begin{equation}
\label{eq:rm}
r_m=1+0.001158(1)C_FN_fg^4\,.
\end{equation}
From our study we will see
that the non-perturbative values are much larger.

Differences of AWI quark masses are related by the non-singlet renormalization
constant $Z$ of Eq.~\eqref{eq:perturb}
to differences of the above lattice quark masses
while the average AWI quark mass $\overline{\widetilde{m}}$ is related by
$Zr_m$ to the average lattice quark mass~\cite{Bhattacharya:2005rb,Sint:1997jx,Sint:1997dj,Fritzsch:2012wq}.
Below we will make use of the relations
\begin{align}
2\left({m}_{13}^2-{m}_{12}^2\right)&=
\frac12\left({m}_s+{m}_{\ell}\right)^2-2{m}_{\ell}^2\nonumber\\
&=-\frac{1}{6}({m}_s-{m}_{\ell})^2+2(m_s-m_{\ell})\overline{m}\,,\\
\overline{m^2}&=\overline{m}^2+\frac{2}{9}(m_s-m_{\ell})^2\,.
\label{eq:msq}
\end{align}
Differences and sums of AWI masses read:
\begin{align}
\widetilde{m}_s-\widetilde{m}_{\ell}
&=2\left(\widetilde{m}_{13}-\widetilde{m}_{12}\right)\nonumber\\
&=Z
\left\{ (m_s-m_{\ell})+
 a\left[ -\frac{\mathcal{A}}{6}\left({m}_s-{m}_{\ell}\right)^2
\right.\right.
\nonumber\\\label{eq:nonsinglet}
 &\qquad+
(2\mathcal{A}+3\mathcal{B})(m_s-m_{\ell})\overline{m}\biggr]\biggr\}\,,\\
\overline{\widetilde{m}}&=\frac13\left(\widetilde{m}_{12}
+\widetilde{m}_{23}+\widetilde{m}_{31}\right)=\frac13\left(\widetilde{m}_s+2\widetilde{m}_{\ell}\right)\nonumber\\
&=Z\left\{r_m\overline{m}+a\left[\frac{1}{18}\left(\mathcal{A}+12\mathcal{D}\right)\left(m_s-m_{\ell}\right)^2\right.\right.\nonumber\\&\qquad+
\left(\mathcal{A}+3\mathcal{B}+9\mathcal{C}+3\mathcal{D}\right)\overline{m}^2
\biggr]\biggr\}\,,\label{eq:nonnonsinglet}
\end{align}
where $\widetilde{m}_{\ell}\equiv\widetilde{m}_{12}$, assuming $\kappa_1=\kappa_2$.
Having rewritten everything in terms of 
differences of quark masses and average quark masses, we can
re-express the improvement terms through AWI masses, which
will allow us to eliminate $\kappa_{\mathrm{crit}}$ from most
equations:
\begin{align}
\widetilde{m}_s-\widetilde{m}_{\ell}&=Z(m_s-m_{\ell})
\nonumber\\
\label{eq:nnsinglet}
&-
\frac{a}{Z}\left[\frac{\mathcal{A}}{6}\left(\widetilde{m}_s-\widetilde{m}_{\ell}\right)^2
+\frac{\mathcal{B}_0}{r_m}\overline{\widetilde{m}}\left(\widetilde{m}_s-\widetilde{m}_{\ell}\right)\right]\,,\\
\overline{\widetilde{m}}&=Zr_m\overline{m}-\frac{a}{Z}\left[\frac{\mathcal{C}_0r_m}{9}\left(\widetilde{m}_s-\widetilde{m}_{\ell}\right)^2
+\frac{\mathcal{D}_0}{2r_m}\overline{\widetilde{m}}^2\right]\,,\label{eq:nsinglet}
\end{align}
where we substituted the combinations $\mathcal{B}, \ldots$ by
$\mathcal{B}_0, \ldots$, that are normalized
such that $\mathcal{A}=\mathcal{B}_0=\mathcal{C}_0=\mathcal{D}_0=1$ at tree-level:
\begin{align}
\mathcal{B}_0&=-2\mathcal{A}-3\mathcal{B}\nonumber\\\label{eq:constants}
&=
-(r_m+1)(b_P-b_A)-2b_m-3(\tilde{b}_P-\tilde{b}_A+\tilde{b}_m)\,,\\
\mathcal{C}_0&=-\frac{1}{2r_m}(\mathcal{A}+12\mathcal{D})=-\frac{1}{2r_m}(b_P-b_A)-2d_m\,,\label{eq:constants3}\\
\mathcal{D}_0&=-\frac{2}{r_m}(\mathcal{A}+3\mathcal{B}+9\mathcal{C}+3\mathcal{D})
\nonumber\\
\label{eq:constants4}
&=-2(b_P-b_A+d_m)-6(\tilde{b}_P-\tilde{b}_A+\tilde{d}_m)\,.
\end{align}
We also made the $r_m$-dependence
explicit since this, though formally $1+\mathcal{O}(g^4)$
[see Eq.~\eqref{eq:rm}], will turn
out significantly larger than one.
Note that $d_m=b_m+\mathcal{O}(g^4)=-1/2+\mathcal{O}(g^2)$
and $b_P=b_A+\mathcal{O}(g^2)=1+\mathcal{O}(g^2)$
[see Eqs.~\eqref{eq:perturb}--\eqref{eq:perturb3}]
while all improvement coefficients $\tilde{d}_J$ and $\tilde{b}_J$
only receive contributions at
$\mathcal{O}(g^4)$.

Although not needed here, for completeness we 
also express lattice quark mass combinations
through AWI masses:
\begin{align}
m_s-m_{\ell}&=\frac{1}{2a}
\left(\frac{1}{\kappa_s}-\frac{1}{\kappa_{\ell}}\right)\nonumber\\
&=\frac{\widetilde{m}_s - \widetilde{m}_{\ell}}{Z}\nonumber\\&\times
\left\{1+a\left[\frac{\mathcal{A}}{6Z}
(\widetilde{m}_s-\widetilde{m}_{\ell})+\frac{\mathcal{B}_0}{Zr_m}\overline{\widetilde{m}}\right]\right\}\,,\\
\overline{m}&=\frac{1}{6a}\left(
\frac{1}{\kappa_s}+\frac{2}{\kappa_{\ell}}-\frac{3}{\kappa_{\mathrm{crit}}}\right)
\nonumber\\\nonumber
&=\frac{\overline{\widetilde{m}}}{Zr_m}\\
&\times\left\{1+
a\left[\frac{\mathcal{C}_0Zr_m}{9Z^2}\frac{\left(\widetilde{m}_s-\widetilde{m}_{\ell}\right)^2}{\overline{\widetilde{m}}}
+\frac{\mathcal{D}_0}{2Zr_m}\overline{\widetilde{m}}\right]\right\}\,.
\end{align}

\subsection{Leading order determination of the target $\kappa_s$ and 
$\kappa_{\ell}$ parameters}
\label{sec:target}
We need to predict how we have to adjust $\kappa_s$ as a function
of $\kappa_{\ell}<\kappa_{\ell,\mathrm{ph}}$ to keep $\widetilde{m}_s=
\widetilde{m}_{s,\mathrm{ph}}$ fixed (and therefore the renormalized
strange quark mass $\widehat{m}_s$ approximately constant).
We assume that the parameters $\kappa_{\ell,\mathrm{ph}}$ and $\kappa_{s,\mathrm{ph}}$
that correspond to the physical point have already been determined, see~Sec.~\ref{sec:mass}.
Neglecting $\mathcal{O}(am)$ terms, we can write
\begin{align}
3\widetilde{m}_s&=2\left(\widetilde{m}_s-\widetilde{m}_{\ell}\right)+
3\overline{\widetilde{m}}\nonumber\\
&=\frac{Z}{2a}\left[2\left(\frac{1}{\kappa_s}-\frac{1}{\kappa_{\ell}}\right)
+r_m\left(\frac{1}{\kappa_s}+\frac{2}{\kappa_{\ell}}-\frac{3}{\kappa_{\mathrm{crit}}}\right)\right]\,,\label{eq:funny}
\end{align}
where we keep the left hand side constant. Solving $\widetilde{m}_s=\widetilde{m}_{s,\mathrm{ph}}$ for
$1/\kappa_s$ gives
\begin{equation}\label{eq:kappas1.2}
\frac{1}{\kappa_s} =  \frac{2}{2+r_m}
\left(\frac{3a}{Z}\widetilde{m}_{s,\mathrm{ph}} +
(1-r_m)\frac{1}{\kappa_{\ell}} + \frac{3 r_m}{2} 
\frac{1}{\kappa_{\mathrm{crit}}} \right)\,.
\end{equation}
$\kappa_{\mathrm{crit}}$ and $\widetilde{m}_{s,\mathrm{ph}}$ can be eliminated by 
subtracting the result at the physical point from
both sides of
this equation:
\begin{equation}\label{eq:kappasf}
 \frac{1}{\kappa_s} =\frac{1}{\kappa_{s,\mathrm{ph}}} +
\frac{2(1-r_m)}{2+r_m} \left(  \frac{1}{\kappa_{\ell}} - \frac{1}{\kappa_{\ell,\mathrm{ph}}} \right)\,,
\end{equation}
i.e.\
for predicting the target $\kappa_s$ as a function of $\kappa_{\ell}$,
up to $\mathcal{O}(a)$ corrections, we
only need to know the value of the parameter
$r_m$, in addition to determining
the physical point. Note that $r_m>1$, which means that
as we decrease $\kappa_{\ell}$ away from the physical point, we have to increase
$\kappa_s$.

The equality
$3\widetilde{m}_{\ell}=(\widetilde{m}_{\ell}-\widetilde{m}_s)+3\overline{\widetilde{m}}$
results in the relation
\begin{equation}
\label{eq:light}
3\widetilde{m}_{\ell}
=\frac{Z}{2a}\left(\frac{r_m-1}{\kappa_s}+\frac{2r_m+1}{\kappa_{\ell}}-\frac{3r_m}{\kappa_{\mathrm{crit}}}\right)\,,
\end{equation}
in analogy to Eq.~\eqref{eq:funny}. We can again
eliminate $\kappa_{\mathrm{crit}}$, by subtracting the result
at the physical point. Furthermore, we substitute
the difference
$\kappa_s^{-1}-\kappa_{s,\mathrm{ph}}^{-1}$ of Eq.~\eqref{eq:kappasf},
giving
\begin{equation}
\label{eq:light2}
a(\widetilde{m}_{\ell}-\widetilde{m}_{\ell,\mathrm{ph}})
=\frac{Z}{2}\frac{3 r_m}{2+r_m}\left(
\frac{1}{\kappa_{\ell}}-\frac{1}{\kappa_{\ell,\mathrm{ph}}}\right)\,.
\end{equation}
If we are now targeting a value 
$\widetilde{m}_{\ell}\neq \widetilde{m}_{\ell,\mathrm{ph}}$,
we can extract the corresponding shift in $\kappa_{\ell}^{-1}$ relative
to the physical point from the above equation.

\subsection{Order-$a$ improvement}
\label{sec:ordera}
Here we show how to order-$a$ improve Eq.~\eqref{eq:kappasf}. However,
we refrain from working out the equivalent expression
for the light quark mass shift Eq.~\eqref{eq:light2} as for our purposes
an approximate determination of the target $\kappa_{\ell}$ value is sufficient.
Above we have not only related AWI masses to lattice masses,
introducing $Z$ and $r_m$, but also improvement terms
with coefficients $\mathcal{A}$, $\mathcal{B}_0$,
$\mathcal{C}_0$ and $\mathcal{D}_0$ have been worked out.

Multiplying Eq.~\eqref{eq:nnsinglet} by two and Eq.~\eqref{eq:nsinglet}
by three,
it is easy to derive an order-$a$ improved version of
Eq.~\eqref{eq:funny}:
\begin{align}
3\widetilde{m}_s&=\frac{Z}{a}\left(\frac{2+r_m}{2\kappa_s}+\frac{r_m-1}{\kappa_{\ell}}-
\frac{3r_m}{2\kappa_{\mathrm{crit}}}\right)\nonumber\\\nonumber
&-\frac{a}{3Z}\biggl[(\mathcal{A}+\mathcal{C}_0r_m)\left(\widetilde{m}_s-\widetilde{m}_{\ell}\right)^2\\&+\left.
\frac{6\mathcal{B}_0}{r_m}\left(\widetilde{m}_s-\widetilde{m}_{\ell}\right)\overline{\widetilde{m}}
+\frac{9\mathcal{D}_0}{2r_m}\overline{\widetilde{m}}^2\right]\,.
\label{eq:morefunny}
\end{align}
Next we keep $\widetilde{m}_s \equiv \widetilde{m}_s(\kappa_{\ell},\kappa_s) = \widetilde{m}_{s,\mathrm{ph}}$ fixed
and compute differences between simulated and physical mass values:
\begin{widetext}\begin{align}
\left(\widetilde{m}_s-\widetilde{m}_{\ell}\right)^2-\left(\widetilde{m}_s-\widetilde{m}_{\ell,\mathrm{ph}}\right)^2&=
\frac{Z}{2a}\left(\frac{1}{\kappa_{\ell}}
-\frac{1}{\kappa_{\ell,\mathrm{ph}}}\right)\left(\widetilde{m}_{\ell}+\widetilde{m}_{\ell,\mathrm{ph}}-2\widetilde{m}_s\right)\,,\\
\left(\widetilde{m}_s-\widetilde{m}_{\ell}\right)\overline{\widetilde{m}}
-\left(\widetilde{m}_s-\widetilde{m}_{\ell,\mathrm{ph}}\right)\overline{\widetilde{m}}_{\mathrm{ph}}&=
\frac{Z}{6a}\left(\frac{1}{\kappa_{\ell}}
-\frac{1}{\kappa_{\ell,\mathrm{ph}}}\right)\left[\widetilde{m}_s
-2\left(\widetilde{m}_{\ell}+\widetilde{m}_{\ell,\mathrm{ph}}\right)\right]\,,\\
\overline{\widetilde{m}}^2-\overline{\widetilde{m}}_{\mathrm{ph}}^2
&=
\frac{2Z}{9a}\left(\frac{1}{\kappa_{\ell}}
-\frac{1}{\kappa_{\ell,\mathrm{ph}}}\right)
\left(\widetilde{m}_s+\widetilde{m}_{\ell}+\widetilde{m}_{\ell,\mathrm{ph}}\right)\,.
\end{align}
\end{widetext}
Above we have re-expressed the non-singlet
combination $\widetilde{m}_{\ell}-\widetilde{m}_{\ell,\mathrm{ph}}$ through
differences of inverse hopping parameters.
We now isolate $1/\kappa_s$ in Eq.~\eqref{eq:morefunny}
and subtract the physical point values from both sides
of the resulting equation:
\begin{widetext}
\begin{align}
 \frac{1}{\kappa_s} -\frac{1}{\kappa_{s,\mathrm{ph}}} = 
\frac{2}{2+r_m}& \left(\frac{1}{\kappa_{\ell}}-\frac{1}{\kappa_{\ell,\mathrm{ph}}}\right)
\biggl\{1-r_m+\frac{a}{6Z}\biggl[(\mathcal{A}+\mathcal{C}_0r_m)
\left(-2\widetilde{m}_{s,\mathrm{ph}}+\widetilde{m}_{\ell,\mathrm{ph}}+\widetilde{m}_{\ell}\right)\nonumber\\\label{eq:improvekappa}
&+\left.\left.\frac{2\mathcal{B}_0}{r_m}
\left(\widetilde{m}_{s,\mathrm{ph}}-2\left(\widetilde{m}_{\ell,\mathrm{ph}}+\widetilde{m}_{\ell}\right)\right)
+\frac{2\mathcal{D}_0}{r_m}
\left(\widetilde{m}_{s,\mathrm{ph}}+\widetilde{m}_{\ell,\mathrm{ph}}+\widetilde{m}_{\ell}\right)
\right]\right\}\,.
\end{align}
Using Eq.~\eqref{eq:light2}, we can re-express
$\widetilde{m}_{\ell}$ above in terms of $\kappa_{\ell}$, $\kappa_{\ell,\mathrm{ph}}$ and
$\widetilde{m}_{\ell,\mathrm{ph}}$.
This gives
\begin{equation}
\label{eq:impkap}
 \frac{1}{\kappa_s} =\frac{1}{\kappa_{s,\mathrm{ph}}} +
\frac{2}{3}x\left\{1-r_m+\frac{1}{3}\left[\frac{\mathcal{B}_0+\mathcal{D}_0}{r_m}-(\mathcal{A}+\mathcal{C}_0r_m)\right]\frac{a\widetilde{m}_{s,\mathrm{ph}}}{Z}
+\frac{1}{3}\left(\mathcal{A}+\mathcal{C}_0r_m+\frac{2\mathcal{D}_0-4\mathcal{B}_0}{r_m}\right)\left(\frac{a\widetilde{m}_{\ell,\mathrm{ph}}}{Z}
+\frac{r_m}{4}x\right)\right\}\,,
\end{equation}
\end{widetext}
where
\begin{equation}
x=\frac{3}{2+r_m}\left(\frac{1}{\kappa_{\ell}}-\frac{1}{\kappa_{\ell,\mathrm{ph}}}\right)\,.
\label{eq:xdef}
\end{equation}
We employ Eq.~\eqref{eq:impkap} to determine the order-$a$ improved
value of $\kappa_s$. The last term of this equation is
numerically subleading since
$\widetilde{m}_{\ell,\mathrm{ph}}\approx 0$ and terms
quadratic in $x$ can be neglected too as long as
$\widetilde{m}_{\ell}\ll\widetilde{m}_s$. Therefore, the dominant
order-$a$ correction
amounts to a constant shift of $\kappa_s^{-1}$, relative to
Eq.~\eqref{eq:kappasf}.
Note that at tree-level
$r_m=\mathcal{A}=\mathcal{B}_0=\mathcal{C}_0=\mathcal{D}_0=1$
and therefore, $\kappa_s=\mathrm{const.}$, as it should be in the
non-interacting case.

\subsection{Comment on additional valence quark flavours}
\label{sec:valence}
We now consider the partially quenched situation, introducing
additional ``charm'' valence quarks of the lattice mass
\begin{equation}
m_c=m_4=m_5=\frac{1}{2a}\left(\frac{1}{\kappa_c}-\frac{1}{\kappa_{\mathrm{crit}}}\right)\,,
\end{equation}
where $\kappa_{\mathrm{crit}}$ is still the hopping parameter
for the $N_f=3$ case at vanishing quark masses. Again, one can obtain the
AWI charm quark mass from Eq.~\eqref{eq:quarkm}. As the charm
quark is quenched, rather than using the currents $\widehat{A}_{\mu}^{4j}$
and $\widehat{P}^{4j}$ with $j\in\{1,2,3\}$, we can also just
compute $\widehat{A}_{\mu}^{45}$ and $\widehat{P}^{45}$, pretending we have
two distinct (but mass-degenerate) charm quark flavours.
In this partially quenched situation, rather than working with
the flavour symmetry group
$\textmd{SU}(3)$, we
have to work with the graded group $\textmd{SU}(5|2)$,
replacing mass traces in Eq.~\eqref{eq:massr3}
by supertraces. The changes can be worked
out easily and reduce to the preceding formulae when replacing
the flavour combination $45$ by $12$ and $m_c$ by $m_{\ell}$.
In particular we have
\begin{align}
\widetilde{m}_c=\frac{Z_P}{Z_A}\widehat{m}_c\left\{
1+a\left[(b_P-b_A)m_c+3(\tilde{b}_P-\tilde{b}_A)\overline{m}\right]\right\}\,,
\end{align}
where
\begin{equation}
\widehat{m}_c=\frac{\partial_4\langle 0|\widehat{A}_4^{45}|\pi^{45}\rangle}
{2\langle 0|\widehat{P}^{45}|\pi^{45}\rangle}\,.
\end{equation}
Note that $\overline{m}$ above still denotes the mass average over the
three sea quark flavours only.

\begin{table*}
\caption{\label{tab:parameters}Analysed CLS (and RQCD) ensembles at
$\beta=3.4$ and $\beta=3.55$: mass plane trajectory
($\overline{m}=m_{\mathrm{symm}}$, $m_s=m_{\ell}$ or
$\widetilde{m}_s=\widetilde{m}_{s,\mathrm{ph}}$), ensemble name, 
hopping parameter values, linear extent in terms of the inverse pion mass,
number of lattice points,
$\sqrt{8t_0}/a$, estimates of pion and kaon
masses from assigning $\sqrt{8t_{0,\mathrm{ph}}}=0.4144(70)\,\textmd{fm}$~\cite{Borsanyi:2012zs}
(also the lattice spacings $a$ are estimated in this way),
the AWI quark masses
in lattice units and the number of molecular dynamics units
$N_{\mathrm{MD}}$. In most cases the trajectory length is two
and measurements are taken every four units. We complement
the $\sqrt{8t_0}/a$ values determined in Ref.~\cite{Bruno:2014jqa}
by preliminary estimates obtained on the newly generated ensembles.
Note that the ensembles H101, H200 and N202 are at the same time on
the $\overline{m}=m_{\mathrm{symm}}$ and $m_s=m_{\ell}$ lines.
All D100 results are very preliminary due to insufficient statistics.
We exclude D100, H200 and rqcd019 from further analysis. Preliminary results are typeset in {\it Italics}.
}
\begin{center}
{\footnotesize
\begin{ruledtabular}
\begin{tabular}{ccllcr@{ $\times$}llccllr}
trajectory&ensemble&\multicolumn{1}{c}{$\kappa_{\ell}$}&\multicolumn{1}{c}{$\kappa_s$}&$LM_{\pi}$&$N_t$ & $N_s^3$&$\sqrt{8t_0}/a$&$\frac{M_{\pi}}{\textmd{MeV}}$&
$\frac{M_K}{\textmd{MeV}}$&$a\widetilde{m}_{\ell}$&$a\widetilde{m}_s$&$N_{\mathrm{MD}}$\\\hline
\multicolumn{11}{c}{$\beta=3.4$ [$a= 0.0854(15)\,\textmd{fm}$]}\\\hline
\multirow{5}{*}{$\overline{m}=m_{\mathrm{symm}}$}
  &H101   &0.13675962 &0.13675962    & 5.8 & $96$ & $32^3$  &     4.772(5) & 422 & 422 &     0.009201(39) &     0.009201(39) &  8000\\
  &H102   &0.136865   &0.136549339   & 4.9 & $96$ & $32^3$  &     4.800(6) & 356 & 442 &     0.006502(46) &     0.013836(43) &  7988\\
  &H105   &0.13697    &0.13634079    & 3.9 & $96$ & $32^3$  &     4.819(6) & 282 & 467 &     0.003951(52) &     0.018672(52) & 11332\\
  &C101   &0.13703    &0.136222041   & 4.6 & $96$ & $48^3$  &     4.824(4) & 223 & 476 &     0.002466(32) &     0.021242(34) &  6208\\
  &D100   &0.13709    &0.136103607   & \it 3.2 & $128$ & $64^3$ & \it 4.860(4) & \it 129 & \it 482 & \it 0.000801(30) & \it 0.023552(43) &   492\\\hline
\multirow{3}{*}{$m_s=m_{\ell}$}
  &rqcd019&0.1366     &0.1366        &8.4 & $32$ & $32^3$&\it 4.454(5)&607&607&0.018094(77)&0.018094(77)&1686\\
  &rqcd021&0.136813   &0.136813      &4.7 & $32$ & $32^3$&\it 4.925(12)&340&340&0.005983(63)&0.005983(63)&1541\\
  &rqcd017&0.136865   &0.136865      &3.3 & $32$ & $32^3$&\it 5.100(7)&238&238&0.002799(92)&0.002799(92)&1849\\\hline
\multirow{3}{*}{$\widetilde{m}_s=\widetilde{m}_{s,\mathrm{ph}}$}
  &H107&0.136945665908&0.136203165143 & 5.1 & $96$ & $32^3$&\it 4.665(6)&368&549&0.006662(50)&0.023981(60)&6256\\
  &H106&0.137015570024&0.136148704478 & 3.8 & $96$ & $32^3$&\it 4.751(6)&272&519&0.003775(70)&0.024029(68)&6212\\
  &C102&0.1370508458  &0.136129062556 & 4.6 & $96$ & $48^3$&\it 4.790(4)&223&504&0.002467(34)&0.023956(54)&6000\\\hline
\multicolumn{11}{c}{$\beta=3.55$ [$a= 0.0644(11)\,\textmd{fm}$]}\\\hline
\multirow{5}{*}{$\overline{m}=m_{\mathrm{symm}}$}
  & H200 & 0.137    & 0.137       & 4.4& $96$ & $32^3$ &     6.419(14) & 418 & 418 & 0.006865(22) & 0.006865(22) & 8000\\
  & N202 & 0.137    & 0.137       & 6.4& $128$ & $48^3$& \it 6.427(6)  & 410 & 410 & 0.006854(16) & 0.006854(16) & 3536\\
  & N203 & 0.13708  & 0.136840284 & 5.4& $128$ & $48^3$& \it 6.416(4)  & 345 & 441 & 0.004738(15) & 0.011047(12) & 6172\\
  & N200 & 0.13714  & 0.13672086  & 4.4& $128$ & $48^3$&     6.424(5)  & 283 & 461 & 0.003164(12) & 0.014132(11) & 6800\\
  & D200 & 0.1372   & 0.136601748 & 4.2& $128$ & $64^3$&     6.430(4)  & 199 & 479 & 0.001538(10) & 0.017229(12) & 4000\\\hline
\multirow{3}{*}{$m_s=m_{\ell}$}
  & B250 & 0.1367   & 0.1367   & 7.4& $64$ & $32^3$ & \it 5.873(8)  & 706 & 706 & 0.018772(39) & 0.018772(39) & 1776\\
  & X250 & 0.13705  & 0.13705  & 5.4& $64$ & $48^3$ & \it 6.500(8)  & 347 & 347 & 0.004899(21) & 0.004899(21) & 1380\\
  & X251 & 0.1371   & 0.1371   & 4.2& $64$ & $48^3$ & \it 6.623(9)  & 268 & 268 & 0.002895(25) & 0.002895(25) & 1384\\\hline
\multirow{3}{*}{$\widetilde{m}_s=\widetilde{m}_{s,\mathrm{ph}}$}
  & N204 & 0.137112   & 0.136575049 & 5.5& $128$ & $48^3$& \it 6.290(5) & 352 & 545 & 0.004822(14) & 0.018927(15) & 3692\\
  & N201 & 0.13715968 & 0.136561319 & 4.5& $128$ & $48^3$& \it 6.351(4) & 284 & 522 & 0.003146(14) & 0.018849(15) & 6000\\
  & D201 & 0.137207   & 0.136546436 & 4.1& $128$ & $64^3$& \it 6.409(4) & 198 & 499 & 0.001552(16) & 0.018874(17) & 4312
\end{tabular}
\end{ruledtabular}
}\end{center}
\end{table*}

Expressing the renormalized charm quark mass through lattice quark masses,
we obtain the following relation between the AWI charm mass and
partially quenched lattice quark masses:
\begin{align}
\widetilde{m}_c&=Z\biggl\{
m_c+(r_m-1)\overline{m}
+a\biggl[
\left(\mathcal{A}+3b_m\right)m_c^2\nonumber\\
&\quad\qquad-\left(\mathcal{B}_0+2\mathcal{A}+6b_m\right)\overline{m}m_c
\nonumber\\\nonumber
&\quad\qquad-\frac{1}{18}\left(2r_m\mathcal{C}_0+\mathcal{A}+6b_m\right)
\left(m_s-m_{\ell}\right)^2\\
&\quad\qquad-\frac{1}{2}\left(r_m\mathcal{D}_0-2\mathcal{A}-2\mathcal{B}_0-6b_m\right)
\overline{m}^2
\biggr]\biggr\}\,.
\label{eq:charmawi}
\end{align}
In spite of the fact that the charm quark is quenched,
$r_m$ still appears in the above equation since we defined the
lattice quark mass
relative to the inverse sea quark critical hopping parameter.\footnote{
Note that when setting $\widetilde{m}_c=\widetilde{m}_{\ell}$ and
$m_c=m_{\ell}$, substituting
$m_{\ell}^2=\overline{m}^2-\frac23\overline{m}(m_s-m_{\ell})
+\frac19(m_s-m_{\ell})^2$
and $\overline{m}m_{\ell}=\overline{m}^2-\frac13\overline{m}(m_s-m_{\ell})$,
Eq.~\eqref{eq:charmawi} reduces to the unquenched case
discussed previously, as it should.}

It is now clear how to keep the AWI (or the
renormalized) charm quark mass approximately constant:
\begin{align}
\widetilde{m}_c-\overline{\widetilde{m}}&=Z(m_c-\overline{m})\left[1+\mathcal{O}(a)\right]\,.
\end{align}
In the situation $\overline{m}=m_{\mathrm{symm}}$, unsurprisingly, at least
up to order-$a$ effects, $m_c$ (and therefore $\kappa_c$)
should be kept constant.
One may wonder if, owing to the heavy charm quark mass,
discretization effects may
be substantial. Examining Eq.~\eqref{eq:charmawi} we notice that
as long as $\overline{m}$ is kept constant only the parametrically
small second last term ($m_s-m_{\ell}\ll m_{c,\mathrm{ph}}$) changes.
Therefore, along the $\overline{m}=m_{\mathrm{symm}}$ trajectory,
also keeping track
of the dominant order-$a$ effects, $m_c$, i.e.\ $\kappa_c$,
should remain constant.

In Secs.~\ref{sec:target} and \ref{sec:ordera} we worked out how
$\kappa_s$ has to be varied to keep $\widetilde{m}_s=\widetilde{m}_{s,\mathrm{ph}}$
constant. What happens along this line?
In this case
\begin{equation}
\widetilde{m}_{c,\mathrm{ph}}-\widetilde{m}_{s,\mathrm{ph}}=Z(m_c-m_{s})\left[1+\mathcal{O}(a)\right]
\end{equation}
should remain constant.
This means that in this situation the difference
between $1/\kappa_c$ and $1/\kappa_s$ should be kept approximately fixed, too:
\begin{equation}
\frac{1}{\kappa_c}=\frac{1}{\kappa_s}+
\frac{1}{\kappa_{c,\mathrm{ph}}}-
\frac{1}{\kappa_{s,\mathrm{ph}}}\,.
\end{equation}
The order-$a$ contributions are again captured by
Eq.~\eqref{eq:charmawi} and
the compensation term
can be worked out if needed,
in analogy to the discussion of Sec.~\ref{sec:ordera}
above.

\section{Simulation parameters and fit procedure}
\label{sec:implement}

In this article we present results obtained at two $\beta$ values:
$\beta=3.4$, corresponding to $a\approx 0.085\,\textmd{fm}$,
and $\beta=3.55$ ($a\approx 0.064\,\textmd{fm}$). Investigations
at further lattice spacings are in progress.
At each lattice spacing the simulations
cover four mass points ranging from a pion mass
$M_{\pi}\approx 200\,\textmd{MeV}$
up to $M_{\pi}\approx 420\,\textmd{MeV}$
along the $\overline{m}=m_{\mathrm{symm}}=\mathrm{const}.$
line with $M_{\pi}L>4$ (3.9 in one case).
At $\beta=3.4$  one additional
point exists along this line at $M_{\pi}\approx 129\,\textmd{MeV}$ (D100).
However, as in this case we have only
limited statistics and $M_{\pi}L<4$,
we will discard this ensemble from any fit.
For an overview of the analysed ensembles,
see Table~\ref{tab:parameters}. The lattice spacing and
meson mass estimates are based on the continuum limit value 
of the scale~\cite{Luscher:2010iy}
$\sqrt{8t_0}=0.4144(59)(37)\,\textmd{fm}$ that was
obtained by the BMW Collaboration~\cite{Borsanyi:2012zs}.
We utilize
open boundary conditions in time~\cite{Luscher:2011kk}, with the exception of
the rqcd017, rqcd019, rqcd021, B250, X250 and X251 ensembles, which are
periodic in time for gauge fields and antiperiodic for fermions.
Note that the ensembles under investigation correspond to lattice spacings
at which topological freezing is not yet a major
problem~\cite{Bruno:2014jqa}, enabling us to use (anti)periodic
boundary conditions.
We remark that at $\beta=3.55$ and
$\kappa_{\ell}=\kappa_s=0.137$ there exist two ensembles,
H200 and N202, both with $LM_{\pi}>4$. However, in the first case
$L\approx 2\,\textmd{fm}$ is rather small in physical units.
While this does not appear to affect the AWI mass,
the measured pion mass on the larger volume comes out somewhat lighter.
Therefore, we discard H200 from further analysis.

In addition to the already existing
$\overline{m}=m_{\mathrm{symm}}$ point at $m_s=m_{\ell}$,
we first generated three other points along the flavour symmetric line for both
lattice spacings, one at a bigger and two at smaller values of the quark mass.
The data along these two lines in the quark mass plane
enabled us to estimate $Z$, $r_m$ and $\kappa_{\mathrm{crit}}$
from fits of the form [see Eqs.~\eqref{eq:nnsinglet} and \eqref{eq:nsinglet}]:
\begin{align}
a\widetilde{m}_s-a\widetilde{m}_{\ell}&=\frac{Z}{2}\left(\frac{1}{\kappa_s}-
\frac{1}{\kappa_{\ell}}\right)\nonumber\\
&\times\left[1-\frac{\mathcal{A}}{12}\left(
\frac{1}{\kappa_s}-\frac{1}{\kappa_{\ell}}\right)
-\mathcal{B}_0a\overline{m}\right]\,,
\label{eq:fitdiff}
\end{align}
\begin{equation}
a\overline{\widetilde{m}}=Zr_m\left[a\overline{m}-\frac{\mathcal{C}_0}{36}
\left(
\frac{1}{\kappa_s}-\frac{1}{\kappa_{\ell}}\right)^2-\frac{\mathcal{D}_0}{2}(a\overline{m})^2\right]\,.
\label{eq:fitsum}
\end{equation}
The combination $a\overline{m}=(2\kappa_{\ell}^{-1}+\kappa_s^{-1}-3\kappa_{\mathrm{crit}}^{-1})/6$ above depends on $\kappa_{\mathrm{crit}}$.
To this order in $a$ this parameter can in principle be substituted by
the measured value of $a\overline{\widetilde{m}}/(Zr_m)$ if desired.
This is possible because
we will neither need the value of $\kappa_{\mathrm{crit}}$ for our
determination of the physical point nor for predicting
the $\kappa_s$ trajectory (as a function of $\kappa_{\ell}$) along
which $\widetilde{m}_s\approx (Z_P/Z_A)\widehat{m}_s$ is kept fixed,
see Eqs.~\eqref{eq:kappasf} and \eqref{eq:impkap}.

We were unable to reliably fit seven (or six)
parameters to data from seven ensembles, covering just the
$\overline{m}=m_{\mathrm{symm}}$ and $m_{s}=m_{\ell}$ lines.
Moreover, $\mathcal{D}_0$ is insensitive
to the $\overline{m}=m_{\mathrm{symm}}$ points, $\mathcal{A}$
does not depend on the $m_s=m_{\ell}$ points and
a determination of $\mathcal{B}_0$ requires points at additional positions
in the quark mass plane.
Therefore, we initially made use of the one-loop estimates
\begin{align}
\mathcal{A}  &=1+0.1153(2)C_Fg^2\,,\\
\mathcal{B}_0&=\mathcal{D}_0=1+0.1126(3)C_Fg^2\,, \\
\mathcal{C}_0&=1+0.1140(1)C_Fg^2\,,
\end{align}
see Eqs.~\eqref{eq:adef}, \eqref{eq:perturb1}--\eqref{eq:perturb3}
and \eqref{eq:constants}--\eqref{eq:constants4}.
After investigating various fits with different combinations
of $\mathcal{A}$, $\mathcal{C}_0$ and $\mathcal{D}_0$ as free parameters,
we found that fixing $\mathcal{A}$, $\mathcal{B}_0$ and $\mathcal{D}_0$
to the above one-loop estimates but allowing $\mathcal{C}_0$ to float
gave a good and stable description of the data.
Note that the non-perturbatively determined values of $\mathcal{A}$~\cite{Korcyl:2016ugy}
were not available at the beginning of this study.
This then enabled us to predict the $\kappa_s$ values that corresponded
to our target AWI strange quark mass (see Sec.~\ref{sec:mass} below), using
Eq.~\eqref{eq:impkap}. Table~\ref{tab:parameters}
demonstrates that indeed we managed to keep
the AWI strange quark mass $\widetilde{m}_s$ constant in simulations
with Wilson fermions within statistical errors of 0.3\% and
0.1\% at $\beta=3.4$ and $\beta=3.55$, respectively, see also
Fig.~\ref{fig:strange} below.

\begin{table*}
\caption{\label{tab:fitresults}
Results of global fits to our AWI quark mass
data according to Eqs.~\eqref{eq:fitdiff} and \eqref{eq:fitsum}.
The $\mathcal{A}$ values were determined in Ref.~\cite{Korcyl:2016ugy}. In
the cases where
varying $\mathcal{A}$ within its uncertainty had an effect, a second
error is given to reflect the associated systematics.}
\begin{center}
\begin{ruledtabular}
\begin{tabular}{ccccccccc}
$\beta$ & $\chi^2/N_{\mathrm{DF}}$&$Z$&$r_m$&$\kappa_{\mathrm{crit}}$&$\mathcal{A}$ (no fit)&$\mathcal{B}_0$&$\mathcal{C}_0$&$\mathcal{D}_0$\\\hline
3.4 &32.1/9&0.8710(30)(10)&2.635(94)(5)&0.1369115(27)(1)&2.91(33)&$-$1.55(76)(1)&3.43(30)&10.0(9.1)(0.3)\\
3.55&26.2/10&0.9841(25)(3)&1.530(14)(1)&0.1371718(10)&2.27(14)&$-$0.81(45)(1)&1.89(25)(1)&1.2(1.2)
\end{tabular}
\end{ruledtabular}
\end{center}
\end{table*}
After simulating three additional points along the
$\widetilde{m}_s=\widetilde{m}_{s,\mathrm{ph}}=\mathrm{const}.$ trajectory for
each lattice spacing, we used
the non-perturbative values
$\mathcal{A}= 2.91(33)$ and
$\mathcal{A}= 2.27(14)$ for $\beta=3.4$ and $\beta=3.55$, respectively,
that were obtained employing
coordinate space methods~\cite{Korcyl:2016ugy}, to determine
the remaining six parameters from a combined correlated fit to all
data. Varying $\mathcal{A}$
within its uncertainty~\cite{Korcyl:2016ugy}
had only a very insignificant impact on the remaining six
fit parameters. After excluding the rather heavy (in lattice units)
$\beta=3.4$ rqcd019 point 
from the fit, we have nine and ten
ensembles with 15 and 16 different quark mass values
at our disposal at $\beta=3.4$ and $\beta=3.55$, respectively.

The resulting fit parameters are shown in Table~\ref{tab:fitresults}.
The data only mildly constrain the parameter $\mathcal{D}_0$,
which comes out to be compatible with zero within
large errors. $r_m$, $\mathcal{B}_0$ and $\mathcal{C}_0$ deviate
substantially from the perturbative expectations and
$\mathcal{B}_0$ even comes out negative. However,
as one would expect, $r_m$ as well as
the improvement parameters are closer to 
unity at the larger $\beta$ value. The fits
are discussed in more detail in Secs.~\ref{sec:fit1} and \ref{sec:fit2}
below. Repeating
the $\beta=3.4$ fit, excluding the lightest $m_s=m_{\ell}$ data point
(rqcd017) as this was obtained on a small volume
$LM_{\pi}=3.2<4$, did not significantly impact on any of the fit parameters
but resulted in increased errors in some cases:
$r_m=2.50(20)$, $\kappa_{\mathrm{crit}}=0.1369159(66)$
and $\mathcal{D}_0=-0.7 \pm 17.0$.
Note that the central value of $\mathcal{D}_0$ moved down by
one standard deviation while the errors on $r_m$ and
$\kappa_{\mathrm{crit}}$ approximately doubled.
Including the heavy rqcd019 point hardly affected
any of the fit parameters, with the exception of
$\mathcal{D}_0=16.5\pm 4.4$, indicating a large positive value
of this parameter. However, in view of the rather large
improvement coefficients at $\beta=3.4$, the rqcd019 results
may very well be polluted by significant $\mathcal{O}(a^2)$ effects, which is
why we chose to discard this point from further analysis.

\begin{figure*}
\includegraphics[width=0.49\textwidth]{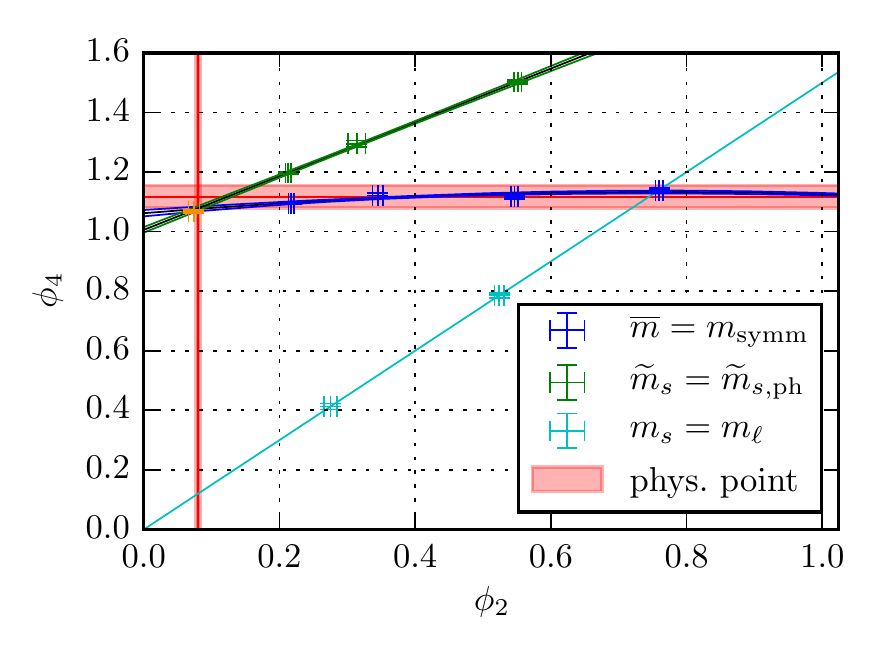}~
\includegraphics[width=0.49\textwidth]{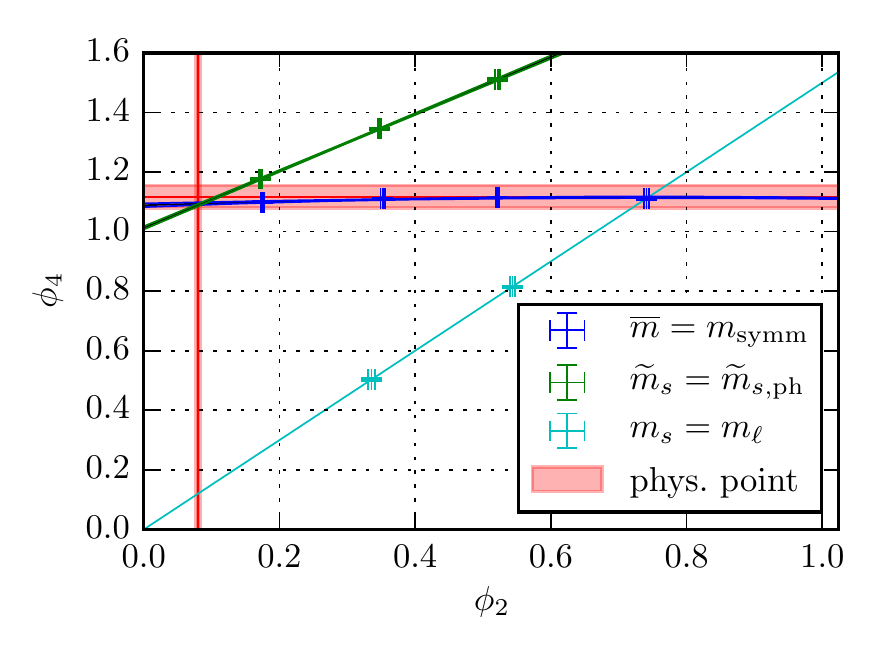}
\caption{
Our simulation points in the $\phi_4\sim \overline{m}$ [see Eq.~\eqref{eq:phi4}]
vs.\ $\phi_2\sim m_{\ell}$ [see Eq.~\eqref{eq:phi21}] plane. The bands correspond
to the uncertainties of the physical point
values Eqs.~\eqref{eq:phi5} and \eqref{eq:phi2}.
$\phi_4=(3/2)\phi_2$ along the $m_s=m_{\ell}$ line and the
other curves represent fits to the data, including (tiny) error bands.
Orange point: ensemble D100. The data points corresponding to ensembles rqcd019 and B250 are outside of the plotted range.
Left: $\beta=3.4$. Right: $\beta=3.55$.}
\label{fig:phiplane}
\end{figure*}

\section{Determination of the physical quark mass values}
\label{sec:mass}
At each lattice spacing the average lattice quark mass
$m_{\mathrm{symm}}$ for the $\overline{m}=m_{\mathrm{symm}}$ trajectory
is fixed by imposing a target value for the
combination~\cite{Bruno:2014jqa}
\begin{align}
\label{eq:phi4}
\phi_4&\equiv 8t_0\left(M_K^2+\frac12 M_{\pi}^2\right)\propto\overline{\widehat{m}}\\\nonumber
&\propto\overline{m}+a\left[\frac{2d_m}{9}(m_s-m_{\ell})^2+(d_m+3\tilde{d}_m)\overline{m}^2\right]
+\mathcal{O}(a^2)
\end{align}
at $\kappa_{\ell}=\kappa_s=\kappa_{\mathrm{symm}}$,
where $t_0$ is the gluonic scale defined in Ref.~\cite{Luscher:2010iy}
and we have made use of the Gell-Mann--Oakes--Renner relation
as well as of Eqs.~\eqref{eq:massr3} and \eqref{eq:msq}.
At the physical point the numerical value
\begin{equation}
\label{eq:phi5}
\phi_{4,\mathrm{ph}}=1.117(38)
\end{equation}
can be obtained from the pion and kaon masses in
the electrically neutral
isospin symmetric limit, $M_{\pi}=134.8(3)\,\textmd{MeV}$ and
$M_K=494.2(4)\,\textmd{MeV}$~\cite{Aoki:2013ldr}, and
the result $\sqrt{8t_0}=0.4144(59)(37)\,\textmd{fm}$
of the BMW Collaboration~\cite{Borsanyi:2012zs} for
the continuum limit $N_f=2+1$ theory.
In the future we will independently determine
a lattice scale and at that stage the value Eq.~\eqref{eq:phi5}
may change.

The combination
$\phi_4$ will not vary strongly along the
$\overline{m}=m_{\mathrm{symm}}$ trajectory: At fixed renormalized
quark masses $\phi_4$ (and $\phi_2$ defined below) can only be subject
to $\mathcal{O}(a^2)$ lattice artefacts. This also holds when
$\overline{\widehat{m}}$ is varied as the effect of
the renormalization of the charge through $b_g$ cancels
from this combination. However, the proportionality of $\phi_4$ to the
average lattice quark mass $\overline{m}$
is subject to order-$a$ corrections,
see Eqs.~\eqref{eq:massr3} and \eqref{eq:phi4}. The latter of the
correction terms in Eq.~\eqref{eq:phi4}
does not change along the $\overline{m}=m_{\mathrm{symm}}$ line.
The remaining $\mathcal{O}(a)$ correction term is proportional to
$(m_s-m_{\ell})^2$.
On the continuum side, generalizing the
Ademollo--Gatto theorem~\cite{Ademollo:1964sr} and also as a consequence
of the Gell-Mann--Okubo
expansion~\cite{GellMann:1962xb,Okubo:1961jc,Bietenholz:2011qq}, $\phi_4$
cannot depend linearly on the symmetry breaking parameter
$m_s-m_{\ell}$, i.e.\ it can only depend on $(M_K^2-M_{\pi}^2)^2$ and
higher powers.
Indeed, to next-to-leading order $\textmd{SU}(3)$
ChPT~\cite{Gasser:1984gg,Bar:2013ora} $\phi_4$ is constant as long as
$\overline{\widehat{m}}$ is constant and
it will only receive corrections at next-to-next-to-leading
order~\cite{Amoros:1999dp}.
Therefore, the $\overline{m}=m_{\mathrm{symm}}$ line with
$\phi_{4,\mathrm{symm}}=\phi_{4,\mathrm{ph}}$, where
$\phi_{4,\mathrm{symm}}$ refers to the $\phi_4$  value at the
point $m_s=m_{\ell}$, should go
through the physical point, up to continuum and lattice effects
that are both quartic in the pseudoscalar
meson masses in units of the chiral symmetry breaking scale
$4\pi F_0\sim 1\,\textmd{GeV}$. In view of these corrections
that depend on $(M_K^2-M_{\pi}^2)^2$ and $\phi_4$ itself,
we targeted a slightly larger value $\phi_{4,\mathrm{symm}}
\approx 1.15>\phi_{4,\mathrm{ph}}$, see Ref.~\cite{Bruno:2014jqa}.

\begin{figure*}
\includegraphics[width=0.49\textwidth]{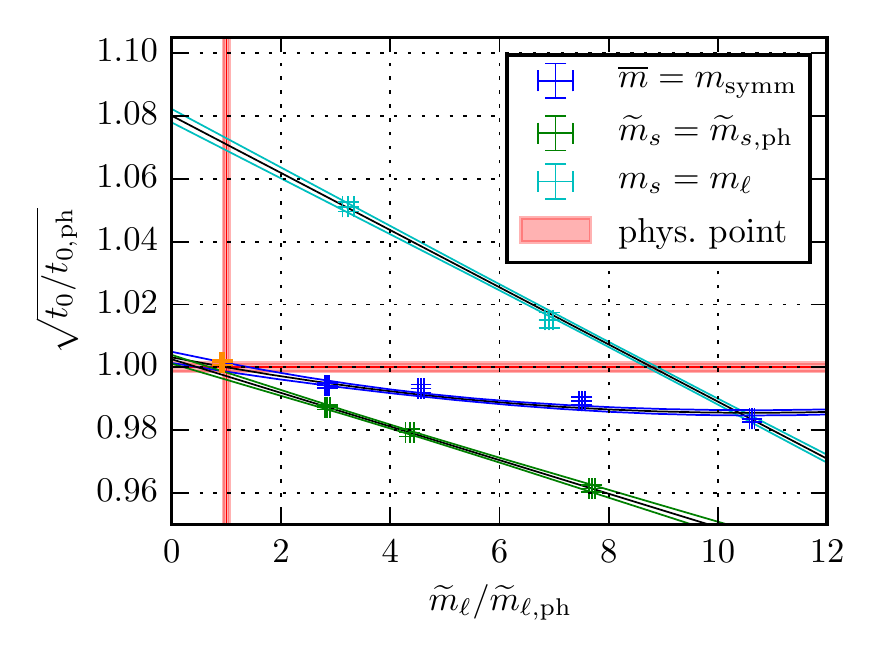}~
\includegraphics[width=0.49\textwidth]{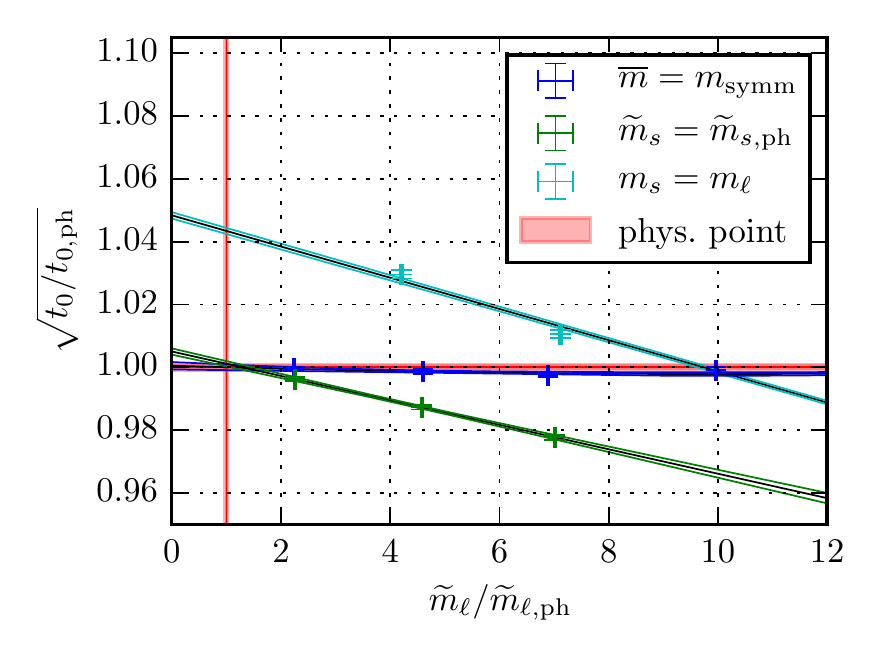}
\caption{
$\sqrt{t_0}$ as a function of the AWI
quark mass $\widetilde{m}_{\ell}$, normalized with respect to
the extrapolated physical point values $\sqrt{t_{0,\mathrm{ph}}}$
and $\widetilde{m}_{\ell,\mathrm{ph}}$,
for our three quark mass plane trajectories.
The lines are fits to the data. Orange point: ensemble D100.
The data points corresponding to ensembles rqcd019 and B250
are outside of the plotted range.
Left: $\beta=3.4$. Right: $\beta=3.55$.
\label{fig:t0}}
\end{figure*}

In the overview plot Fig.~\ref{fig:phiplane} the positions of our analysed
ensembles in the $\phi_4$ vs.\ 
\begin{equation}
\label{eq:phi21}
\phi_2\equiv 8t_0M_{\pi}^2\propto \widehat{m}_{\ell}
\end{equation}
plane are shown, where at the physical point:
\begin{equation}
\label{eq:phi2}
\phi_{2,\mathrm{ph}}=0.0801(28)\,,
\end{equation}
using again the physical values for $t_0$ and $M_\pi$ from above.
Abscissa and ordinate are approximately proportional to the light and
average lattice quark masses, respectively.
The physical target ranges $\phi_{4,\mathrm{ph}}$ and $\phi_{2,\mathrm{ph}}$
are shown
as horizontal and vertical error bands.
The line for $m_s=m_{\ell}$ corresponds to $\phi_4=3\phi_2/2$,
for $\widetilde{m}_s=\widetilde{m}_{s,\mathrm{ph}}$ we show a linear
fit to the data while for $\overline{m}=m_{\mathrm{symm}}$
we plot a constant plus quadratic function of
$\phi_{2,\mathrm{symm}}
-\phi_2\propto t_0(M_K^2-M_{\pi}^2)$.
Indeed, the $\phi_4$ combination only mildly varies
between the $\overline{m}=m_{\mathrm{symm}}$ simulation points
but changes significantly along the other two mass plane trajectories.
The orange point at $\beta=3.4$ corresponds to the ensemble
D100 (see Table~\ref{tab:parameters}) that does not enter
any of our fits.

Along the $\overline{m}=m_{\mathrm{symm}}$ trajectory
order-$a$ lattice artefacts as well as the leading continuum
chiral correction both are proportional to $(m_s-m_{\ell})^2$.
Because of this variation the target value
at the $\textmd{SU}(3)$ symmetric point was deliberately
chosen larger than its physical point estimate Eq.~\eqref{eq:phi5}.
At both couplings we have somewhat undershot this
target value $\phi_{4,\mathrm{symm}}=1.15$:
For our $\beta=3.4$ and $\beta=3.55$ data we obtain
$\phi_{4,\mathrm{symm}}=1.139(6)$ and $\phi_{4,\mathrm{symm}}=1.111(4)$,
respectively. This then results in smaller
than physical central $\phi_{4,\mathrm{ph}}$ values.
From the fit to the $\overline{m}=m_{\mathrm{symm}}$ points we
find
\begin{align}
\frac{\phi_{4,\mathrm{ph}}}{\phi_{4,\mathrm{symm}}}&=0.945(9)\,,
&\phi_{4,\mathrm{ph}}=1.076(8)\,,\\
\frac{\phi_{4,\mathrm{ph}}}{\phi_{4,\mathrm{symm}}}&=0.985(6)\,,
&\phi_{4,\mathrm{ph}}=1.094(5)\,,
\end{align}
at $\beta=3.4$ and $\beta=3.55$, respectively.
In both cases the slope of $\phi_4$ as a function of
$(\phi_{2,\mathrm{symm}}-\phi_2)^2$
is small but significant. It decreases towards the smaller lattice
spacing, indicating that the main effect may be due to
lattice artefacts.
In hindsight, if we would have chosen $\phi_{4,\mathrm{symm}}\approx 1.18$
and $\phi_{4,\mathrm{symm}}\approx 1.13$ at $\beta=3.4$ and $\beta=3.55$,
respectively, we would have hit the central physical target
value for $\phi_4$ at $\phi_2=\phi_{2,\mathrm{ph}}$
along our $\overline{m}=m_{\mathrm{symm}}$ line.
Nevertheless, both extrapolated $\phi_{4,\mathrm{ph}}$ values
agree with the physical value
$\phi_{4,\mathrm{ph}}=1.117(38)$ within errors.
We remark that also this value is not final
but depends on a future independent determination of the scale parameter
$t_0$.

In Fig.~\ref{fig:t0} we plot the ratio $\sqrt{t_0/t_{0,\mathrm{ph}}}$
as a function of the light AWI quark mass ratio
$\widetilde{m}_{\ell}/\widetilde{m}_{\ell,\mathrm{ph}}$ where we normalize
with respect to the corresponding physical mean values (see below).
At both lattice spacings $\sqrt{t_0}$ along the
$\overline{m}=m_{\mathrm{symm}}$ line depends only mildly
on $m_{\ell}$ but --- as expected --- it
changes considerably along the other lines, and in particular
along $m_s=m_{\ell}$. The dependence on
the quark masses can be parameterized
in terms of lattice spacing and continuum
effects~\cite{Bar:2013ora}. The latter are to leading order
linear functions of
$\overline{\widetilde{m}}\propto\overline{\widehat{m}}$
and $(\widetilde{m}_s-\widetilde{m}_{\ell})^2
\propto (\widehat{m}_s-\widehat{m}_{\ell})^2$, where the
proportionalities hold up to tiny residual $\mathcal{O}(a)$
effects (see above). Extrapolating the $\overline{m}=m_{\mathrm{symm}}$
data quadratically in $\widetilde{m}_s-\widetilde{m}_{\ell}
\propto \overline{\widetilde{m}}-\widetilde{m}_{\ell}$,
we obtain the physical point values
\begin{align}
\label{eq:t01}
\frac{\sqrt{t_{0,\mathrm{ph}}}}{\sqrt{t_{0,\mathrm{symm}}}}&=1.0167(18)\,,
&\frac{\sqrt{8t_{0,\mathrm{ph}}}}{a}=4.852(7)\,,\\
\frac{\sqrt{t_{0,\mathrm{ph}}}}
{\sqrt{t_{0,\mathrm{symm}}}}&=1.0009(13)\,,
&\frac{\sqrt{8t_{0,\mathrm{ph}}}}{a}=6.433(6)\,,
\label{eq:t02}
\end{align}
for $\beta=3.4$ and $\beta=3.55$, respectively.
Note that some of the $t_0$ values shown in
the figure are preliminary, see Table~\ref{tab:parameters},
and may have underestimated errors, due to very large
autocorrelations times for this observable, that we may not yet have
taken fully into account. Therefore, the
$\sqrt{8t_{0,\mathrm{ph}}}/a$ values quoted above should also be considered
as preliminary.
The curves shown in the figure are linear fits to the
$\widetilde{m}_s=\widetilde{m}_{s,\mathrm{ph}}$ and
$m_s=m_{\ell}$ data and the constant plus quadratic fit
in $\overline{\widetilde{m}}-\widetilde{m}_{\ell}$ to the
$\overline{m}=m_{\mathrm{symm}}$ data described above.

Note that linear lattice
artefacts along the $m=m_s=m_{\ell}=\overline{m}$ line 
are due to the change of the
coupling $\tilde{g}^2=(1+b_gam)g^2$.
This means that in this case the dependence on the lattice spacing $a$
of the linear slope of $\sqrt{t_0(\widehat{m})}$ as a function of $m$ 
could serve to isolate
the effect of $b_g$, when compared to the continuum
limit mass dependence $t_{0,\mathrm{c}}(\widehat{m})$ along this line:
From 
\begin{align}
a(\tilde{g}^2)&=
\Lambda^{-1}\exp\left(-\frac{8\pi^2}{\beta_0\tilde{g}^2}+\cdots\right)\nonumber\\
&=a(g^2)\left(1+b_aam+\cdots\right)
\end{align}
it follows that
$b_a=8\pi^2b_g/\beta_0=8\pi^2b_g/9$ for $N_f=3$. This coefficient
is related to the change of the slope
\begin{equation}
\frac{\sqrt{t_0(\widehat{m})}}{\sqrt{t_0(0)}}
=\frac{a(g^2)}{a(\tilde{g}^2)}
\frac{\sqrt{t_{0,\mathrm{c}}(\widehat{m})}}{\sqrt{t_{0,\mathrm{c}}(0)}}
=(1-b_aam)\frac{\sqrt{t_{0,\mathrm{c}}(\widehat{m})}}{\sqrt{t_{0,\mathrm{c}}(0)}}\,.
\end{equation}
Note that $d_m + 3\tilde{d}_m$ and $Z_mr_m$ are required to relate
$\sqrt{t_0(m)}$ to $\sqrt{t_0(\widehat{m})}$.
As the slope becomes more negative
for the coarser lattice spacing, $b_g$ must be positive,
in agreement with the one-loop perturbative expectation~\cite{Luscher:1996sc}:
A larger average quark mass results in a coarser effective
lattice spacing.

\begin{figure*}
\includegraphics[width=0.49\textwidth]{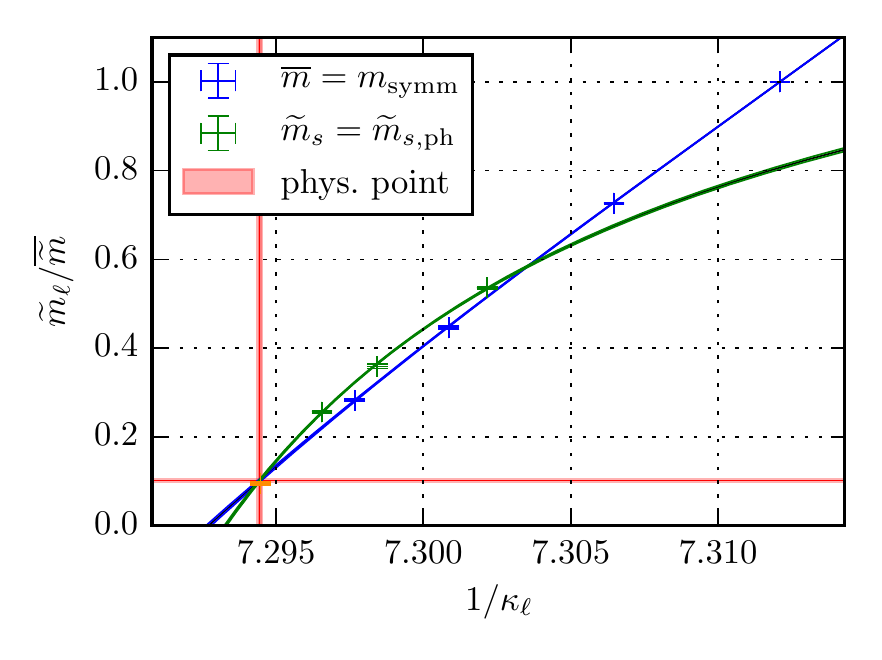}~
\includegraphics[width=0.49\textwidth]{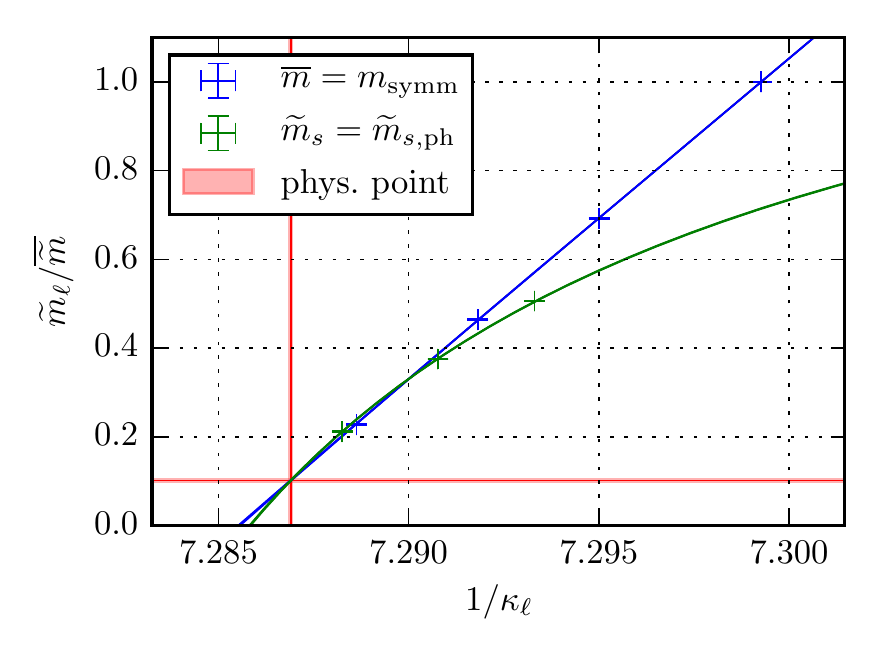}
\caption{
Extrapolation of the AWI
mass ratio $\widetilde{m}_{\ell}/\overline{\widetilde{m}}$ of Eq.~\eqref{eq:target}
to the physical point (horizontal and vertical bands).
The curves correspond to
Eqs.~\eqref{eq:fitdiff} and \eqref{eq:fitsum} with the parameter values of
Table~\protect\ref{tab:fitresults}. Orange point: ensemble D100.
Left: $\beta=3.4$. Right: $\beta=3.55$.\label{fig:massdef}}
\end{figure*}

As AWI masses can be determined more precisely than pseudoscalar
masses and are less susceptible to finite volume effects, we
will use these to set the
physical quark masses, by imposing the FLAG value~\cite{Aoki:2013ldr},
\begin{equation}
\label{eq:target}
\frac{\widetilde{m}_{\ell,\mathrm{ph}}}
{\overline{\widetilde{m}}_{\mathrm{ph}}}\approx
\frac{\widehat{m}_{\ell,\mathrm{ph}}}
{\overline{\widehat{m}}_{\mathrm{ph}}}=0.1018(15)\,,
\end{equation}
to define the physical quark mass point.
Also this target value may undergo a slight
change in the future,
once we have independently extrapolated the combination
\begin{equation}
\label{eq:piextra}
\frac{3M_{\pi}^2}{2M_{K}^2+M_{\pi}^2}=0.1076(5)
\end{equation}
to the continuum limit.

We remark that there are slight differences between ratios of
renormalized and AWI quark masses
$\widehat{m}$ and $\widetilde{m}$, see Eq.~\eqref{eq:hatmass}. Since
we keep $\overline{m}$ fixed along
our main mass plane trajectory,
the dependence on $\tilde{b}_P-\tilde{b}_A$ cancels from the ratio
Eq.~\eqref{eq:target}. However, a correction term
\begin{equation}
a\frac{b_P-b_A}{3}(m_{\ell,\mathrm{ph}}-m_{s,\mathrm{ph}})=
a\frac{b_P-b_A}{3Z}(\widetilde{m}_{\ell,\mathrm{ph}}-\widetilde{m}_{s,\mathrm{ph}})
\end{equation}
survives, see Eq.~\eqref{eq:hatmass}. 
Non-perturbatively, one finds~\cite{Korcyl:2016ugy}
$b_P-b_A= 0.90(32)$ and $b_P-b_A=0.59(14)$, respectively,
at $\beta=3.4$ and $\beta=3.55$. At $\beta=3.4$, where the above
order-$a$ correction is largest, using $Z^{-1}\approx 1.15$ and
$a(\widetilde{m}_{s,\mathrm{ph}}-\widetilde{m}_{\ell,\mathrm{ph}})\approx 0.023$
(see Table~\ref{tab:fitresults} and Eq.~\eqref{eq:tar34}), we obtain a change of about
0.8\% from substituting the ratio of renormalized quark masses
Eq.~\eqref{eq:target} by the ratio of AWI masses.
This effect, that reduces to 0.4\% at $\beta=3.55$,
is well below the 1.5\% relative error of the FLAG average~\cite{Aoki:2013ldr}
that we use.

The global fit of
Eqs.~\eqref{eq:fitdiff} and \eqref{eq:fitsum} to our mass data
provides a parametrization of the AWI masses as
functions of $\kappa_{\ell}$ and $\kappa_s$. The fit parameters
can be found in Table~\ref{tab:fitresults} and the fit is discussed
in detail in Secs.~\ref{sec:fit1} and \ref{sec:fit2} below.
At each lattice spacing we then determine the physical hopping parameter values
$\kappa_{\ell,\mathrm{ph}}$ and $\kappa_{s,\mathrm{ph}}$ as well as the corresponding
AWI masses $\widetilde{m}_{\ell,\mathrm{ph}}$ and $\widetilde{m}_{s,\mathrm{ph}}$
that satisfy Eq.~\eqref{eq:target} along
the chiral trajectory $\overline{m}=m_{\mathrm{symm}}$.
Note that since we have a parametrization of light and
strange AWI masses as functions of the hopping parameters,
we can also determine the physical point along any
other chiral trajectory that incorporates it.\footnote{In the absence
of an independent determination of the scale parameter $t_0$,
for the moment being we fix $\frac13\sum_i\kappa_i^{-1}=0.13675962$ and
$\frac13\sum_i\kappa_i^{-1}=0.137$ at $\beta=3.4$ and $\beta=3.55$,
respectively, i.e.\ we assume that our $\overline{m}=m_{\mathrm{symm}}$
curves go exactly through the physical point.
As can be seen in Fig.\ref{fig:phiplane} this assumption is justified.
Along this line the physical point is then defined by Eq.~\eqref{eq:target}.}
Our procedure of finding the physical point is illustrated in Fig~\ref{fig:massdef}, where we plot this ratio as
a function of $\kappa_{\ell}^{-1}$. In addition to the
$\overline{m}=m_{\mathrm{symm}}$ points (blue) that follow lines
with little curvature, as expected from
Eqs.~\eqref{eq:fitdiff} and \eqref{eq:fitsum}, we also show the results
of our subsequent measurements along
the $\widetilde{m}_s=\widetilde{m}_{s,\mathrm{ph}}$
trajectory (green), which nicely coincide with the parametrization,
thereby validating our strategy.
Note that the $m_s=m_{\ell}$ points  (with the exception
of the symmetric point
on the $\overline{m}=m_{\mathrm{symm}}$ line) are not shown
as the ratio displayed is trivial in this case. The error band
of the target value is dominated by the uncertainty of
Eq.~\eqref{eq:target}.

\begin{figure}
\includegraphics[width=0.5\textwidth]{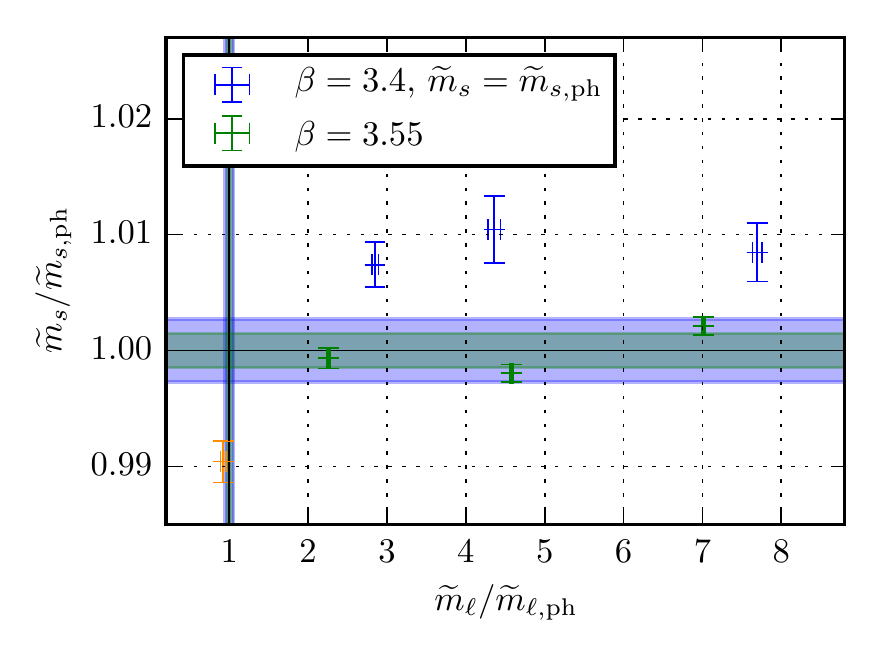}
\caption{
The strange quark AWI masses along the
$\widetilde{m}_s=\widetilde{m}_{s,\mathrm{ph}}$ curves, normalized
with respect to the (postdicted) central
physical point values Eqs.~\eqref{eq:tar34} and \eqref{eq:tar35},
as a function of the corresponding light quark mass ratios.
The error bands are dominated by the uncertainty of
the target value Eq.~\eqref{eq:target}, where we neglected
the uncertainty of $m_{\mathrm{symm}}$. Orange point: $\beta=3.4$
ensemble D100.
\label{fig:strange}}
\end{figure}

The physical point values, postdicted 
including more statistics and the newly generated fixed AWI strange quark
mass ensembles, read:
\begin{align}
&\kappa_{\ell,\mathrm{ph}}=0.1370906(13)\,,
&\kappa_{s,\mathrm{ph}}=0.1361024(25)\,,\\
&a\widetilde{m}_{\ell,\mathrm{ph}}=0.000866(48)\,,
&a\widetilde{m}_{s,\mathrm{ph}}=0.023780(65)\,,\label{eq:tar34}
\end{align}
and
\begin{align}
&\kappa_{\ell,\mathrm{ph}}=0.1372326(5)\,,
&\kappa_{s,\mathrm{ph}}=0.1365373(10)\,,\\
&a\widetilde{m}_{\ell,\mathrm{ph}}=0.000688(17)\,,
&a\widetilde{m}_{s,\mathrm{ph}}=0.018887(28)\,,\label{eq:tar35}
\end{align}
at $\beta=3.4$ and $\beta=3.55$, respectively.
Our original targets had been
$a\widetilde{m}_{s,\mathrm{ph}}=0.0240$ at $\beta=3.4$ and
$a\widetilde{m}_{s,\mathrm{ph}}=0.0189$ at $\beta=3.55$.
In the latter case this agrees with the corresponding value
above. At $\beta=3.4$, however, we mistuned by 1\%
since the estimate of $c_A$~\cite{Bulava:2013afa,Bulava:2015bxa}
changed during our study.

In Fig.~\ref{fig:strange} we plot
our measured AWI strange quark masses determined
on the newly generated ensembles along the predicted
$\widetilde{m}_s=\widetilde{m}_{s,\mathrm{ph}}$ trajectory,
normalized with respect to the postdicted values from the global fit
shown in Eqs.~\eqref{eq:tar34} and \eqref{eq:tar35} above as
a function of the light quark AWI mass. Indeed,
the two sets of AWI strange quark
masses are constant within errors. The $\beta=3.55$ data
perfectly coincide with the expectation
(see also the N204, N201 and D201 entries of
Table~\ref{tab:parameters}).
The $\beta=3.4$ data agree reasonably well with the
original target value
(see ensembles H107, H106 and C102 of Table~\ref{tab:parameters})
but they are off by almost 3 standard deviations, corresponding
to 1\%, from the postdiction Eq.~\eqref{eq:tar34}, that was
obtained using an altered value of $c_A$~\cite{Bulava:2015bxa}.

Neglecting any uncertainty on
$2m_{\ell}+m_s=3m_{\mathrm{symm}}$, the 1.5\% error of the target range
Eq.~\eqref{eq:target} translates into a small error
on $\widetilde{m}_{s,\mathrm{ph}}$ and a larger relative error on
$\widetilde{m}_{\ell,\mathrm{ph}}$. These uncertainties contribute to the
horizontal and vertical error bands shown in the figure.
Note that these would be wider if we could include a realistic error
estimate for $m_{\mathrm{symm}}$.
The figure demonstrates that it is possible to
tune the AWI strange quark mass to the desired value within a few
per mille and even in the case where, due to the
incomplete information on $c_A$, we mistuned to an
incorrect target value the difference
is extremely small.

\section{Discussion of the fits and the resulting parameters}
\label{sec:test}
We will first investigate in more detail lattice spacing effects
for the $\overline{m}=m_{\mathrm{symm}}$ and $m_s=m_{\ell}$ data,
before presenting an overview of all AWI mass data, and discussing
combinations of improvement coefficients.
\subsection{\boldmath${\overline{m}=m_{\mathrm{symm}}}$ and $m_s=m_{\ell}$ data and
improvement coefficients}
\label{sec:fit1}
\begin{figure}
\includegraphics[width=0.5\textwidth]{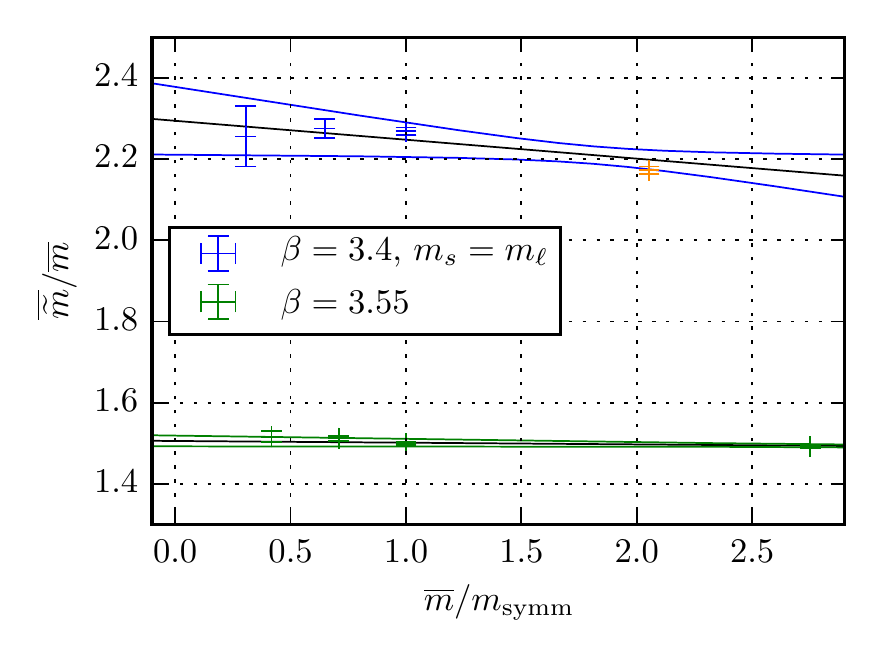}
\caption{
The ratio of the AWI over the
lattice quark mass along the $m_s=m_{\ell}$ trajectory, as
a function of $\overline{m}=m_s=m_{\ell}$, normalized to
the mass $m_{\mathrm{symm}}\approx \overline{m}_{\mathrm{ph}}$.
The value of $\overline{\widetilde{m}}/\overline{m}$
at $\overline{m}=0$ corresponds to the combination of renormalization
constants $Zr_m$. Orange point: ensemble rqcd019 (excluded from the global
fit). The curves correspond to
Eq.~\eqref{eq:msymm}, with the parameter
values of Table~\ref{tab:fitresults}.
\label{fig:msym}}
\end{figure}
Here we compare different projections of our quark mass data
to the global fits
Eqs.~\eqref{eq:fitdiff} and \eqref{eq:fitsum} with the parameter
values shown in Table~\ref{tab:fitresults}.

In Fig.~\ref{fig:msym}
we show the ratio of the
AWI over the
lattice quark mass $\overline{\widetilde{m}}/\overline{m}=\widetilde{m}/m$
as a function of $\overline{m}$
for our $m_s=m_{\ell}=m=\overline{m}$ data.
Note that the right-most orange $\beta=3.4$ point (rqcd019)
did not enter the fit as we suspect this may be polluted
by significant $\mathcal{O}(a^2)$ effects.
To enable a direct comparison
between different lattice spacings, the $x$-axis is normalized
with respect to $m_{\mathrm{symm}}$, the average
lattice quark mass used along our
$\overline{m}=m_{\mathrm{symm}}$ trajectory.
From Eq.~\eqref{eq:fitsum} we can see that
\begin{equation}
\label{eq:msymm}
\frac{\overline{\widetilde{m}}}{\overline{m}}=
Zr_m\left\{1-a\left[\frac{\mathcal{C}_0}{9}
\frac{\left(m_s-m_{\ell}\right)^2}{\overline{m}}+\frac
{\mathcal{D}_0}{2}\overline{m}\right]\right\}\,.
\end{equation}
Therefore, we can directly read off the combination $Zr_m$ at
$\overline{m}=0$ from the figure (as well as from Table~\ref{tab:fitresults}).
Since the difference
$m_s-m_{\ell}$ vanishes for the
data shown, the slope corresponds to the combination
$-Zr_m\mathcal{D}_0am_{\mathrm{symm}}/2$. This becomes better
constrained towards the larger $\beta$ value but --- as its
effect is small --- the deviation of this parameter
from the tree-level expectation $\mathcal{D}_0=1$ is
hard to extract. This also means that our results are
quite insensitive regarding the value of $\mathcal{D}_0$.

\begin{figure}
\includegraphics[width=0.5\textwidth]{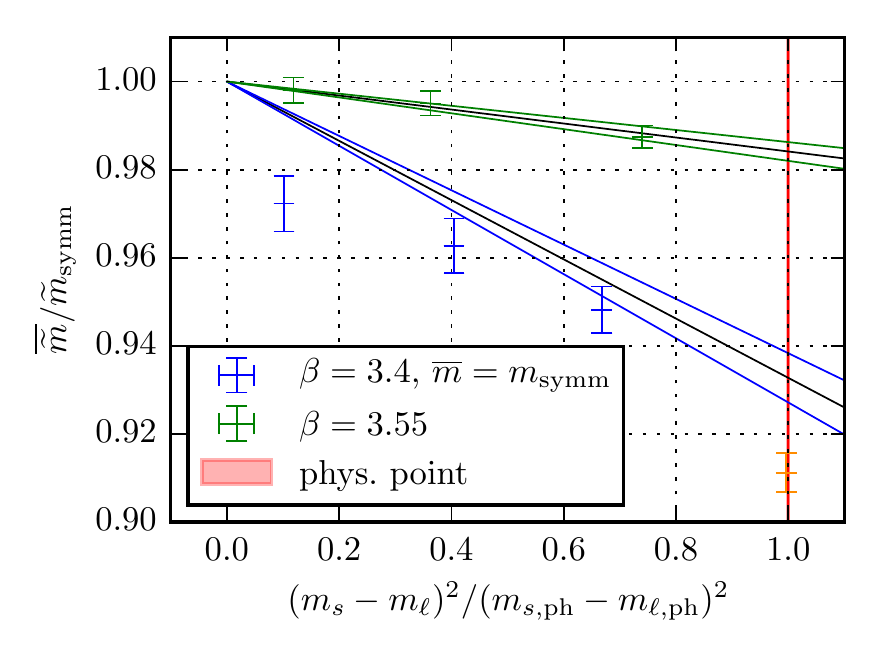}
\caption{
The dependence of the average AWI quark mass
on the difference of the lattice quark masses squared, along
the $\overline{m}=m_{\mathrm{symm}}$ trajectory. Orange point:
ensemble D100. Both
mass combinations are normalized with respect to
their $m_s=m_{\ell}$ value $\widetilde{m}_{\mathrm{symm}}$
and their physical point value, respectively.
The curves correspond to
Eq.~\eqref{eq:mmmmmm} with the parameter
values of Table~\ref{tab:fitresults}.}
\label{fig:mbar}
\end{figure}

\begin{figure}
\includegraphics[width=0.5\textwidth]{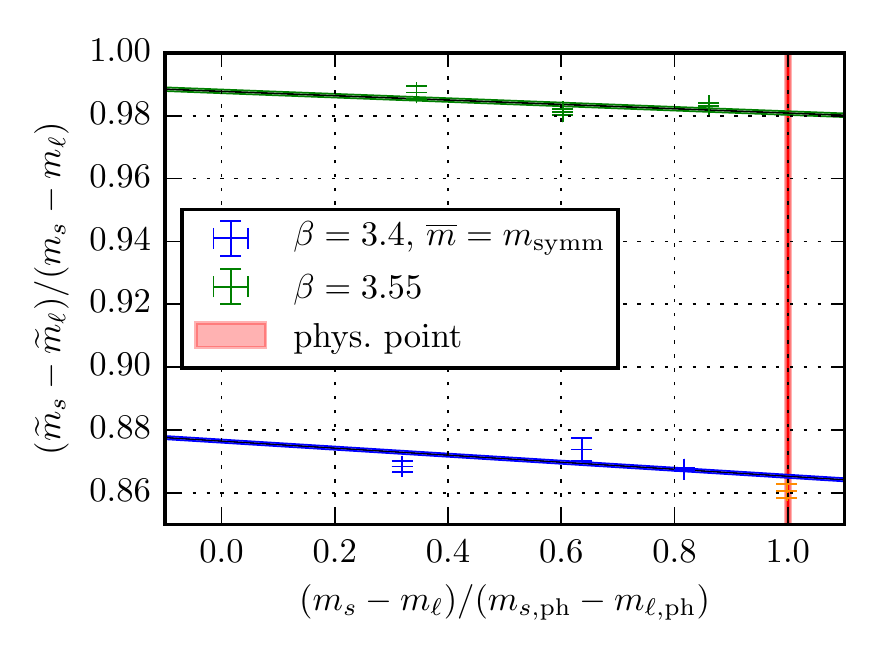}
\caption{The ratio between AWI and
lattice quark mass differences along the $\overline{m}=m_{\mathrm{symm}}$ trajectory, as
a function of $m_s-m_{\ell}$, normalized to
this mass difference at the physical point. Orange point:
ensemble D100. Up to an order-$a$ term,
the value of the ratio at $m_s-m_{\ell}=0$ corresponds to
the renormalization constant combination $Z$.
The curves correspond to
Eq.~\eqref{eq:m2345} with the parameter
values of Table~\ref{tab:fitresults}.}
\label{fig:mdif}
\end{figure}

\begin{figure*}
\includegraphics[width=0.49\textwidth]{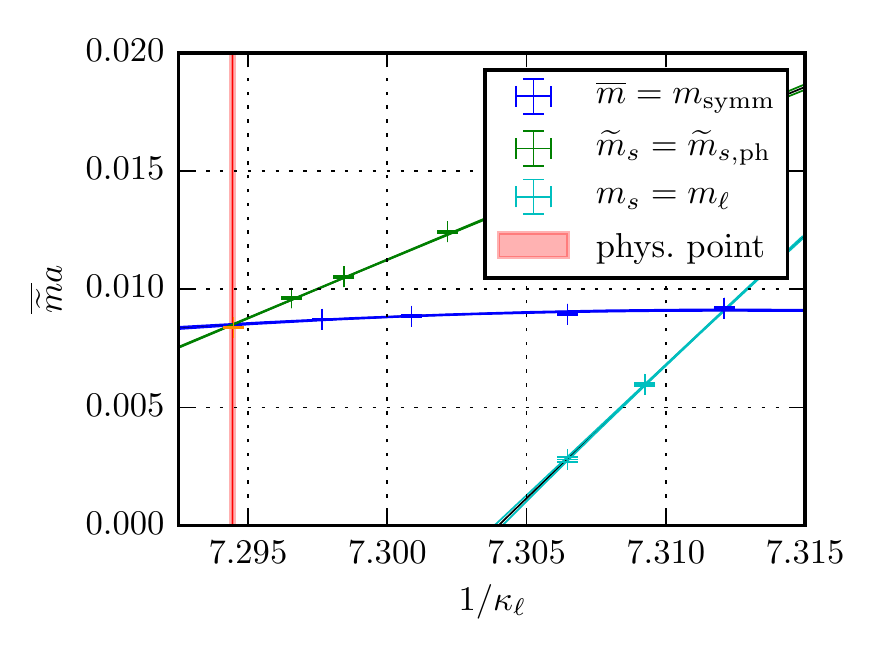}~
\includegraphics[width=0.49\textwidth]{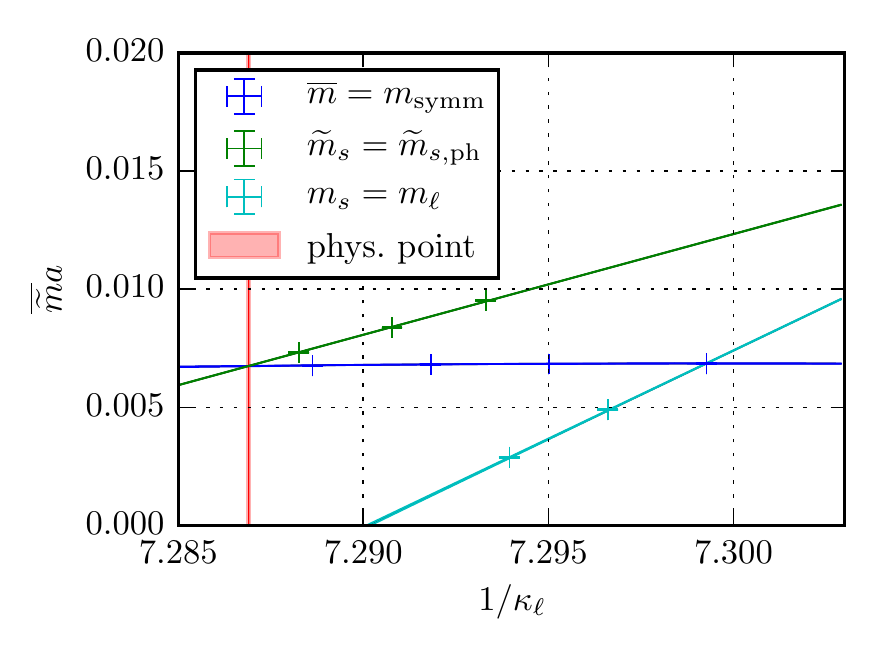}
\caption{
The average AWI quark mass along our three trajectories,
together with the fit according to Eq.~\eqref{eq:fitsum}
with the parameter values of Table~\ref{tab:fitresults}
vs.\ $\kappa_\ell^{-1}$. Orange point: ensemble D100.
Left: $\beta=3.4$. Right: $\beta=3.55$.\label{fig:mbarb}}
\end{figure*}

In Fig.~\ref{fig:mbar} we show a combination that
isolates the effect of $\mathcal{C}_0$: $\overline{\widetilde{m}}$
over $\widetilde{m}_{\mathrm{symm}}$ as a function of
$(m_s-m_{\ell})^2$, normalized to the corresponding physical point value,
for the constant average lattice quark mass data
$\overline{m}=m_{\mathrm{symm}}$. As the light quark mass decreases
the ratio shown deviates from one.
The parametrization can be read off from Eq.~\eqref{eq:msymm}:
\begin{equation}
\label{eq:mmmmmm}
\frac{\overline{\widetilde{m}}}{\widetilde{m}_{\mathrm{symm}}}
=1-\frac{\mathcal{C}_0a(m_{s,ph}-m_{\ell,ph})^2}{9m_{\mathrm{symm}}}
\left(\frac{m_s-m_{\ell}}{m_{s,\mathrm{ph}}-m_{\ell,\mathrm{ph}}}\right)^2.
\end{equation}
This means the negative slope
is proportional to $\mathcal{C}_0$, which decreases considerably,
increasing $\beta$ from 3.4 to 3.55, see also Table~\ref{tab:fitresults}.
For $\beta=3.4$ we show the preliminary D100 
physical point result (orange) that did not enter our global fit.
The left-most point at $\beta=3.4$ (corresponding to the H102 
ensemble of Table~\ref{tab:parameters}) exhibits the
largest deviation of our data from the two global fits to
data from 9 and 10 ensembles (15 and 16 quark mass values),
respectively. Since $a\overline{m}$ is constant
for all the $\beta=3.4$ data shown in the figure and $a(m_s-m_{\ell})$ is
larger for the other points, we would not expect the
average AWI quark mass on ensemble H102 to be
particularly sensitive to higher order discretization effects.
Therefore, we assume the deviation seen is a statistical fluctuation.

\begin{figure*}
\includegraphics[width=0.49\textwidth]{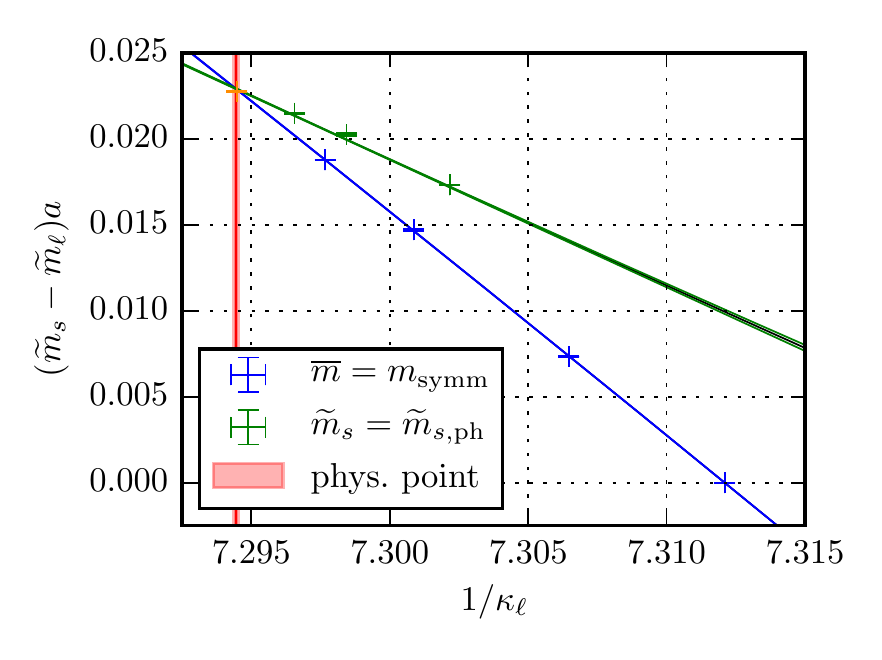}~
\includegraphics[width=0.49\textwidth]{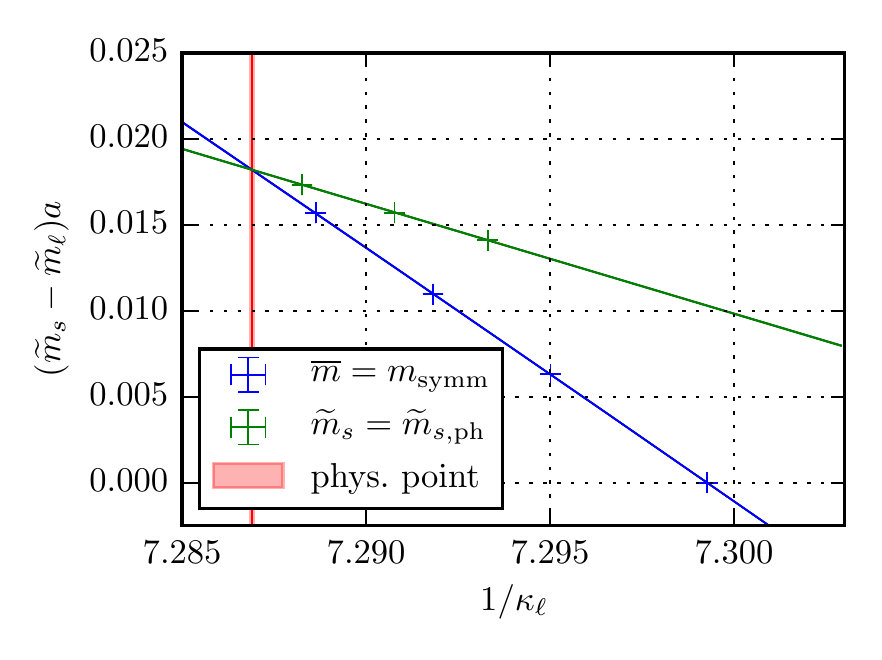}
\caption{Differences between AWI quark masses,
together with the fit according to Eq.~\eqref{eq:fitdiff}
with the parameter values of Table~\ref{tab:fitresults}
vs.\ $\kappa_\ell^{-1}$. Orange point: ensemble D100.
Left: $\beta=3.4$. Right: $\beta=3.55$.\label{fig:dmb}}
\end{figure*}

\begin{figure*}
\includegraphics[width=0.49\textwidth]{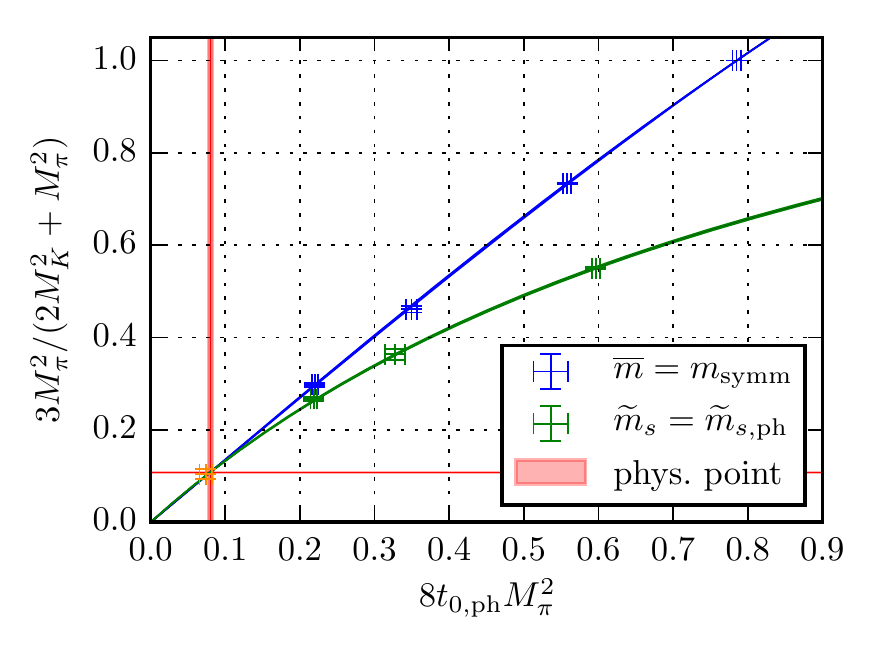}~
\includegraphics[width=0.49\textwidth]{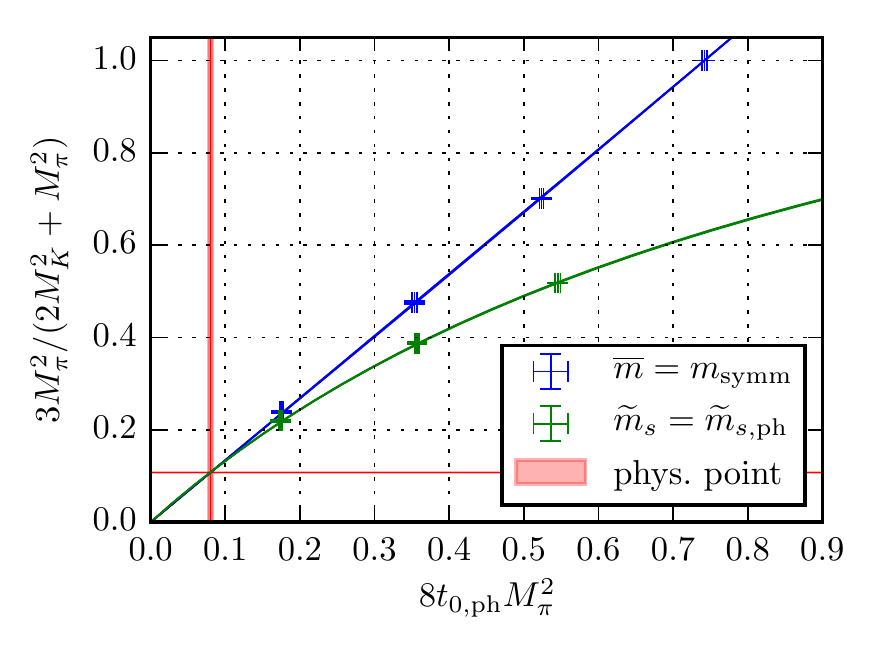}
\caption{
The ratio $3M_{\pi}^2/(2M_K^2+M_{\pi}^2)$ as a function
of $8t_{0,\mathrm{ph}}M_{\pi}^2$ along the $\overline{m}=m_{\mathrm{symm}}$
and $\widetilde{m}_s=\widetilde{m}_{s,\mathrm{ph}}$ mass plane
trajectories for our two lattice spacings. The vertical and
horizontal bands correspond to the physical point.
The parametrization of the fit curves is given in
Eqs.~\eqref{eq:ff1} and \eqref{eq:ff2}. Orange point: ensemble D100.
Left: $\beta=3.4$. Right: $\beta=3.55$.
\label{fig:piextra}}
\end{figure*}

Finally, in Fig.~\ref{fig:mdif} we show the ratio of AWI and
lattice quark mass differences as a function of
$m_s-m_{\ell}$ for the $\overline{m}=m_{\mathrm{symm}}$ data.
We normalize the ordinate with respect
to its physical point value, to enable comparison between
different lattice spacings. We also display 
the global fits. Note that the right-most orange
$\beta=3.4$ point (D100) did not enter the fit as
we regard its error as unreliable, due to limited
statistics. From Eq.~\eqref{eq:fitdiff}
we see that
\begin{equation}
\label{eq:m2345}
\frac{\widetilde{m}_s-\widetilde{m}_{\ell}}{m_s-m_{\ell}}=
Z\left\{1-a\left[\frac{\mathcal{A}}{6}
\left(m_s-m_{\ell}\right)+
\mathcal{B}_0\overline{m}\right]\right\}\,.
\end{equation}
Therefore, at $m_s=m_{\ell}$ we can read off the combination
$Z(1-\mathcal{B}_0am_{\mathrm{symm}})$. This
is somewhat larger than the $Z$ parameters
that are listed in Table~\ref{tab:fitresults} because
$\mathcal{B}_0$ is negative in both cases.
The slope corresponds to
$-Z\mathcal{A}a(m_{s,\mathrm{ph}}-m_{\ell,\mathrm{ph}})/6$.
Note that the value of the parameter $\mathcal{A}$ that was
obtained independently~\cite{Korcyl:2016ugy} is consistent with
our data.

We have demonstrated that $\mathcal{A}$, $\mathcal{C}_0$
and $\mathcal{D}_0$ can be constrained from data along
the $\overline{m}=m_{\mathrm{symm}}$ and $m_s=m_{\ell}$ lines.
While the sensitivity to $\mathcal{D}_0$ is small,
$\mathcal{B}_0$ cannot be constrained at all from data points
along these two lines.
Obviously, $\overline{m}$ needs to be varied for
$m_s\neq m_{\ell}$ for the results to
become sensitive to $\mathcal{B}_0$. 
Unfortunately, to enable a precise determination of $Z$ [and of $r_m=(Zr_m)/Z$]
some knowledge of $\mathcal{B}_0$ is required but we were only
able to constrain
$\mathcal{B}_0$, once data points from other regions of the
quark mass plane were added, in our case along the
$\widetilde{m}_{s}=\widetilde{m}_{s,\mathrm{ph}}$ line.
Another possibility
of achieving this would have been to follow
a partially quenched strategy, computing valence
quark AWI masses along the sea quark symmetric line $m_s=m_{\ell}$.
Nevertheless, in order to determine the $\widetilde{m}_{s}=\widetilde{m}_{s,\mathrm{ph}}$ line
from the two chiral trajectories $\overline{m}=m_{\mathrm{symm}}$ and $m_s=m_{\ell}$
there is no need for a very accurate value of $\mathcal{B}_0$ since the
$\mathcal{B}_0am_{\mathrm{symm}}$ contribution to $Z$ only amounts to 
a 0.6\% correction, even on the coarser $\beta=3.4$ ensemble.

\subsection{AWI and pseudoscalar mass data}
\label{sec:fit2}
In Fig.~\ref{fig:mbarb} we plot the average AWI mass in lattice units
as a function of $\kappa_{\ell}^{-1}$ for our three quark mass
plane trajectories: $\overline{m}=m_{\mathrm{symm}}$,
$\widetilde{m}_s=\widetilde{m}_{s,\mathrm{ph}}$ and $m_s=m_{\ell}$.
The curves correspond to our fit according to Eq.~\eqref{eq:fitsum}
with parameters as shown in Table~\ref{tab:fitresults}.
The $\overline{m}=m_{\mathrm{symm}}$ curve is almost constant
as along this line $\overline{\widetilde{m}}$ only changes
due to an $\mathcal{O}(a)$ term that is parameterized by
$\mathcal{C}_0$, see Fig.~\ref{fig:mbar}.
The curvature is not visible on the scale of Fig.~\ref{fig:mbarb}. 

In Fig.~\ref{fig:dmb} a comparison is shown
between strange and light AWI mass differences
$\widetilde{m}_s-\widetilde{m}_{\ell}$ and the fit
Eq.~\eqref{eq:fitdiff}. In this representation the $m_s=m_{\ell}$
points obviously coincide with zero and are therefore
not shown. Overall, we have good coverage of the quark
mass plane and the data are described reasonably well by the fit.
Like in Fig.~\ref{fig:mbarb}, at both lattice spacings
that we investigated, the $\overline{m}=m_{\mathrm{symm}}$
and $\widetilde{m}_s-\widetilde{m}_{s,\mathrm{ph}}$ curves intersect
very close to (and in statistical agreement with)
the preferred position of the physical point.

Finally, in Fig.~\ref{fig:piextra} we display the ratio
$3M_{\pi}^2/(2M_K^2+M_{\pi}^2)$
as a function of
$8t_{0,\mathrm{ph}}M_{\pi}^2$ for the $\overline{m}=m_{\mathrm{symm}}$ and
$\widetilde{m}_s=\widetilde{m}_{s,\mathrm{ph}}$ trajectories.
Again, $t_{0,\mathrm{ph}}$ denotes the value of this parameter at the
physical point, which is somewhat larger than
its value $t_{0,\mathrm{symm}}$ at the $\textmd{SU}(3)$ flavour symmetric
point, see Fig.~\ref{fig:t0} and Eqs.~\eqref{eq:t01} and \eqref{eq:t02}.
The curves shown correspond
to the parametrizations
\begin{align}
\label{eq:ff1}
\frac{3M_{\pi}^2}{2M_K^2+M_{\pi}^2}&=\frac{3M_{\pi}^2}{
\alpha M_{\pi}^2+(3-\alpha)M_{\pi,\mathrm{symm}}^2}\,,\\
\frac{3M_{\pi}^2}{2M_K^2+M_{\pi}^2}&=\frac{3M_{\pi}^2}{\gamma/(8t_{0,\mathrm{ph}})
+2M_{\pi}^2}\,,
\label{eq:ff2}
\end{align}
for $\overline{m}=m_{\mathrm{symm}}$ and
$\widetilde{m}_s=\widetilde{m}_{s,\mathrm{ph}}$, respectively.
The functional dependencies
enforce the first curve to take the value one at the symmetric
point and both curves to go through zero for $M_{\pi}=0$.
For the above ratio of pseudoscalar masses we expect to find the value
$0.1076(5)$ [the horizontal band, see Eq.~\eqref{eq:piextra}]
at the physical point, which is defined through 
$8t_{0,\mathrm{ph}}M_{\pi,\mathrm{ph}}^2=\phi_{2,\mathrm{ph}}
=0.0801(28)$ [the vertical band,
see Eq.~\eqref{eq:phi2}].

The dimensionless fit parameters can be related
to meson masses at the symmetric and physical points:
\begin{align}
\alpha&=\frac{2M_{K,\mathrm{ph}}^2+M_{\pi,\mathrm{ph}}^2-3M_{\pi,\mathrm{symm}}^2}{
M_{\pi,\mathrm{ph}}^2-M_{\pi,\mathrm{symm}}^2}\,,\\
\gamma&=8t_{0,\mathrm{ph}}(2M_{K,\mathrm{ph}}^2-M_{\pi,\mathrm{ph}}^2)=2(\phi_{4,\mathrm{ph}}-
\phi_{2,\mathrm{ph}})\,.
\end{align}
The fitted parameter values read
\begin{align}
&\alpha=0.235(19)\,,&\gamma=2.054(12)\,,\\
&\alpha=-0.025(15)\,,&\gamma=2.062(6)\,,
\end{align}
for $\beta=3.4$ and $\beta=3.55$, respectively.
Both $\gamma$ values agree well with the physical point
expectation $2(\phi_{4,\mathrm{ph}}-\phi_{2,\mathrm{ph}})=2.075(70)$.
The curvature of the $\overline{m}=m_{\mathrm{symm}}$ line is
of a higher order in ChPT and also
subject to an $\mathcal{O}(a)$ lattice effect.
The change
of the parameter $\alpha$ with $\beta$ indicates that the latter dominates.

At the physical point we obtain the following values
for the ratio Eq.~\eqref{eq:piextra} from the fits
Eqs.~\eqref{eq:ff1}--\eqref{eq:ff2}:
0.1087(6)(37) and 0.1086(6)(35) for the
$\overline{m}=m_{\mathrm{symm}}$ and
$\widetilde{m}_s=\widetilde{m}_{s,\mathrm{ph}}$ data, respectively,
at $\beta=3.4$ and
0.1074(4)(37) and 0.1082(3)(35) at $\beta=3.55$.
The first errors are statistical only while the second
errors given include the propagation of the
uncertainty of
$\phi_{2,\mathrm{ph}}$
that defines the physical point, which is dominated  
by the scale uncertainty of $t_0$. Just considering
the statistical errors and taking the central value
of $\phi_{2,\mathrm{ph}}$ for granted, all ratios agree
reasonably well with the ``experimental''
value 0.1076(5) quoted in Eq.~\eqref{eq:piextra}:
Imposing the continuum FLAG
ratio~\cite{Aoki:2013ldr} Eq.~\eqref{eq:target}
of renormalized quark masses gives the expected result for this
ratio of experimental pseudoscalar masses, also at our two
finite lattice spacings. It is particularly reassuring that
this is the case independent of the quark mass trajectory.
This should allow us to improve on the precision of present
quark mass determinations, once more lattice spacings are
analysed and the continuum limit has been taken.

\subsection{Further combinations of improvement coefficients}
We have determined the combinations of improvement coefficients
Eqs.~\eqref{eq:constants}--\eqref{eq:constants4} as well
as $r_m$, the ratio of the singlet over non-singlet quark mass
renormalization constants, see Table~\ref{tab:fitresults}.
Additional information on
$\mathcal{A}=b_P-b_A-2b_m$, $b_P-b_A$ and $b_m$ exists from
 Ref.~\cite{Korcyl:2016ugy}.
Using these results, we can estimate other improvement coefficients:
From Eq.~\eqref{eq:constants3} we infer that
\begin{equation}
d_m=-\frac{1}{4}\left(2\mathcal{C}_0+\frac{b_P-b_A}{r_m}\right)\,.
\end{equation}
Equation~\eqref{eq:constants} gives
\begin{equation}
\tilde{b}_m+\tilde{b}_P-\tilde{b}_A=-\frac{1}{3}\left[\mathcal{B}_0
+(r_m+1)(b_P-b_A)+2b_m
\right]\,.
\end{equation}
Then from the above and Eq.~\eqref{eq:constants4} we obtain
\begin{align}
\tilde{d}_m-\tilde{b}_m&=\frac{1}{12}\Biggl[
4\mathcal{B}_0+2\mathcal{C}_0-2\mathcal{D}_0\\\nonumber
&\quad +\frac{1+4r_m^2}{r_m}(b_P-b_A)
+8b_m\Biggr]\,.
\end{align}
We collect the resulting estimates in Table~\ref{tab:improv}.
Like $\mathcal{A}$, $\mathcal{B}_0$, $\mathcal{C}_0$,
$\mathcal{D}_0$ and $r_m$ displayed in Table~\ref{tab:fitresults},
also these parameters appear to converge towards the
perturbative expectations as the lattice spacing is reduced.
In particular $d_m\approx b_m$ holds at $\beta=3.55$.

\begin{table}
\caption{\label{tab:improv}Various improvement
coefficient (combinations). $\mathcal{A}$, $\mathcal{B}_0$, $\mathcal{C}_0$
and $\mathcal{D}_0$ are listed in Table~\ref{tab:fitresults}.
The $b_m$ and $b_P-b_A$ values below have been obtained in Ref.~\cite{Korcyl:2016ugy}.
}
\begin{center}
\begin{ruledtabular}
\begin{tabular}{cccc}
coefficient&perturbation theory&$\beta=3.4$&$\beta=3.55$\\\hline
$b_m$&$-1/2-0.0703g^2$&$-$1.04(31)&$-$0.85(14)\\
$d_m$&$-1/2-0.0703g^2$&$-$1.80(16)&$-$1.04(13)\\
$b_P-b_A$&$0.0012g^2$&0.90(32)&0.59(14)\\
$\tilde{b}_m+\tilde{b}_P-\tilde{b}_A$&$\mathcal{O}(g^4)$&0.12(51)&0.34(21)\\
$\tilde{d}_m-\tilde{b}_m$&$\mathcal{O}(g^4)$&$-$1.5(1.5)&$-$0.39(27)
\end{tabular}
\end{ruledtabular}
\end{center}
\end{table}
\section{Summary and outlook}
We outlined a strategy to keep the strange quark
mass, determined through the axial Ward identity, constant
in simulations with $N_f=2+1$ flavours of Wilson fermions, implementing
full order-$a$ improvement, see Eqs.~\eqref{eq:kappasf},
\eqref{eq:light2} and \eqref{eq:impkap}. This was successfully tested to
very high precision at two
lattice spacings, $a\approx 0.085\,\textmd{fm}$ and
$a\approx 0.064\,\textmd{fm}$, see Fig.~\ref{fig:strange}.
We estimated this procedure to
differ, due to as yet only partially known flavour singlet
$\mathcal{O}(a)$ effects,
from keeping the renormalized strange quark mass constant
by less than one percent, even at the coarser lattice spacing.
Furthermore, we worked out how valence quark hopping parameters need
to be adjusted, see Sec.~\ref{sec:valence}. This will
be used in future studies of charm physics.

We computed several combinations of renormalization
constants and order-$a$ improvement coefficients,
see Tables~\ref{tab:fitresults} and \ref{tab:improv}.
We observed that the parameters
that are either related to flavour singlet quark mass
combinations ($\tilde{b}_m$, $\tilde{b}_P$, $\tilde{b}_A$),
flavour singlet currents ($r_m$, $d_m$) or both ($\tilde{d}_m$)
come out very different from the corresponding
perturbative expectations. However, these seem to converge
rapidly in the direction of the tree-level results when increasing
$\beta=6/g^2$ from 3.4 to 3.55.
Non-perturbative order-$a$ improvement of all currents of
interest is ongoing, see, e.g., Refs.~\cite{Bulava:2015bxa,Korcyl:2016ugy}.

In order to set the physical strange quark mass we also
determined the physical point in the $\kappa_s^{-1}$ vs.\ $\kappa_{\ell}^{-1}$
plane.
Its position may still undergo changes in the future, once the
continuum limit has been taken independently, which
then might necessitate a slight
reweighting~\cite{Finkenrath:2013soa,Leder:2015fea} of the strange quark
mass. Having ensembles along three lines in the
quark mass plane ($2m_{\ell}+m_s=3m_{\mathrm{symm}}$,
$\widetilde{m}_s=\widetilde{m}_{s,\mathrm{ph}}$ and $m_s=m_{\ell}$)
enables tests of the convergence
of $\textmd{SU}(2)$ and $\textmd{SU}(3)$ ChPT
and of expansions in the $\textmd{SU}(3)$ symmetry
breaking parameter~\cite{Bietenholz:2011qq} as well as highly constrained
physical point extrapolations. It also allows us to pursue
a non-perturbative renormalization and order-$a$ improvement
programme.

Results on baryon distribution amplitudes
along the $2m_{\ell}+m_s=\mathrm{const.}$ trajectory at
one lattice spacing have already been published~\cite{Bali:2015ykx}.
Further results on distribution amplitudes,
also utilizing the constant strange quark
mass points, are in preparation and an article on
light hadron spectroscopy is forthcoming.

\acknowledgments
This work was supported by the Deutsche Forschungsgemeinschaft Grant
No.\ SFB/TRR 55. The authors gratefully acknowledge the Gauss Centre for
Supercomputing e.V. (\url{http://www.gauss-centre.eu}) for granting
computer time on SuperMUC at Leibniz Supercomputing Centre
(LRZ, \url{http://www.lrz.de}) and JUQUEEN at J\"ulich Supercomputing
Centre (JSC, \url{http://www.fz-juelich.de/ias/jsc}).
GCS is the alliance of the three national supercomputing centres HLRS
(Universit\"at Stuttgart),
JSC  (Forschungszentrum J\"ulich) and LRZ
(Bayerische Akademie der  Wissenschaften),
funded by the German Federal Ministry of Education and Research (BMBF)
and the German State Ministries for Research of Baden-W\"urttemberg (MWK),
Bayern (StMWFK) and Nordrhein-Westfalen (MIWF). 
The authors also gratefully acknowledge computer time provided by
PRACE (Partnership for Advanced Computing in Europe,
\url{http://www.prace-ri.eu}) as part of the project ContQCD.
Additional simulations were performed on the Regensburg iDataCool cluster
and on the SFB/TRR~55 QPACE computer~\cite{Baier:2009yq,Nakamura:2011cd}.
{\sc openQCD}~\cite{ddopenqcd,Luscher:2012av} was used to generate
the main gauge ensembles,
as part of the joint CLS effort~\cite{Bruno:2014jqa}.
Additional $m_s=m_{\ell}$ ensembles were generated on QPACE (using
{\sc BQCD}~\cite{Nakamura:2010qh,Nakamura:2011cd})
and on the Wilson HPC Cluster at IKP Mainz. Two-point functions
were computed using the {\sc Chroma}~\cite{Edwards:2004sx} software package,
along with the locally deflated domain decomposition solver implementation of
{\sc openQCD}~\cite{ddopenqcd}.
We thank Christian Hoelbling for his permission to reproduce a figure
of Ref.~\cite{Hoelbling:2014uea}.
We thank Mattia Bruno, Sara Collins, Piotr Korcyl and Rainer Sommer
for discussions, Tassos Vladikas for useful comments relating to an
earlier draft, Benjamin Gl\"a\ss{}le for software support,
Fabian Hutzler for the generation of some of the two-point functions
and all our other CLS colleagues.
\bibliography{fixeds2}
\end{document}